\documentstyle[12pt]{article}

\begin{document}

\title{Equivalence of the Euclidean and Wightman Field Theories.}
\author{Yu.~M.~Zinoviev\thanks{The research described in this
publication was made possible in part by Grant No. 93--011--147
from the Russian Foundation of Fundamental Researches.} \thanks{e--mail:
zinoviev@qft.mian.su}
\\ Steklov Mathematical Institute,
Vavilov St. 42,\\ Moscow 117966, GSP-1, Russia}

\date{}
\maketitle

\noindent {\bf Abstract} The new inversion formula for the Laplace
transformation of the tempered distributions with supports in the closed
positive semiaxis is obtained. The inverse Laplace transform of the tempered
distribution is defined by means of the limit of the special distribution
constructed from this distribution. The weak spectral condition on the
Euclidean Green's functions requires that some of the limits needed for
the  inversion formula exist for any Euclidean Green's function with
even number of variables. We prove that the initial
Osterwalder--Schrader axioms \cite{os1} and the weak spectral condition
are equivalent with the Wightman axioms.

\section{Introduction}

\noindent In 1973 K.Osterwalder and R.Schrader \cite{os1} claimed to
have found necessary and sufficient conditions under which Euclidean
Green's functions have analytic continuations whose boundary values
define a unique set of Wightman distributions. The pricipal idea of the
Osterwalder--Schrader paper \cite{os1} was to consider the Euclidean
Green's functions to be the distributions. Usually the Euclidean Green's
functions were considered to be the analytic functions. Later R.Schrader
\cite{os2} found the counter example for the crucial lemma of the paper
\cite{os1}. In 1975 K.Osterwalder and R.Schrader proposed additional
"linear growth condition" under which Euclidean Green's functions,
satisfying the Osterwalder--Schrader axioms \cite{os1}, define the
Wightman theory. But these new extended axioms for the Euclidean
Green's functions may be not equivalent with the Wightman axioms.
It is possible to restore the equivalence theorem by adding the new
condition \cite{os2} that the Euclidean Green's functions are the
Laplace transforms of the tempered distributions with supports in the
positive semiaxis with respect to the time variables. The equivalence
theorem becomes trivial \cite{os2}. This new condition contradicts
the principal Osterwalder--Schrader idea to consider the Euclidean
Green's functions to be the distributions and it is not suitable for
applications because it seems difficult to check it up. This paper
is an attempt to understand the mathematical foundation of the
Osterwalder--Schrader results. Our aim is to find the additional
reasonable condition which allows to prove that the extended
Osterwalder--Schrader axioms are equivalent with the Wightman axioms.

One of the Ostewalder--Schrader axioms is the positivity condition.
If we consider the simplest case and neglect the space variables we
can write the positivity condition in the form

\begin{equation}
\label{1.1}
\int_{0}^{\infty } dt \int_{0}^{\infty} ds f(t + s)
\overline{\phi (t)} \phi (s) \geq 0
\end{equation}
Due to \cite[Lemma A]{gla} the positivity condition (\ref{1.1}) for the
distribution $f(t) \in D^{\prime }({\bf R}_{+})$, where ${\bf R}_{+}$ is
the open positive semiaxis, implies the condition in ${\bf R}_{+}$

\begin{equation}
\label{1.2}
\sum_{m,n} a_{m} \overline{a}_{n}  \frac{d^{m + n}f}{dt^{m +n}} (t)
\geq 0
\end{equation}
for all finite sequences of the complex numbers $a_{m}$. Corollary C
from \cite{gla} implies that the distribution
$f(t) \in D^{\prime }({\bf R}_{+})$, satisfying the condition (\ref{1.2})
for all terminating sequences of complex numbers $a_{m}$, is the
restriction to the semiaxis of a function $A(x + iy)$ analytic in the
tube ${\bf R}_{+} + i{\bf R}$. To explain the difficulties which this
way encounters in proving the Osterwalder--Schrader theorem we cite
here an extract from the remarkable paper \cite{gla} : "The Euclidean
Green's functions satisfying the Osterwalder--Schrader postulates
can be shown to be restrictions of the functions analytic in the whole
Wightman causal domain and to satisfy the positivity condition there
in a sence to be presently explained. The author has, however, not been
able to show the tempered growth of those analytic functions near the
real Minkowski space boundary and believes at present that this is
impossible to achieve without further assumptions on the growth
properties of Schwinger functions $s_{n}$ with respect to the index $n$.
This is suggested by the fact that in order to reach the real Minkowski
space by analytic completion for a given $s_{n}$ an infinite number of steps
are required, each of which involves the other functions $s_{m}$ via the
Schwartz inequality with higher and higher values of $m$". Our way of
proving the equivalence theorem doesn't use the analytic functions at all.
Due to the Osterwalder--Schrader idea we consider the Euclidean Green's
functions to be the distributions.

S.Bernstein \cite{ber} called a function exponentially convex if it satisfies
the positivity condition (\ref{1.1}). We shall prove that a tempered
distribution $f(t) \in S^{\prime } ({\bf R}_{+})$ is exponentially
convex iff a tempered distribution
$g(t) = f(- t) \in S^{\prime }({\bf R}_{-})$ is absolutely monotonic, i. e.

\begin{equation}
\label{1.3}
\frac{d^{m}g}{dt^{m}} (t) \geq 0
\end{equation}
for all $m = 0,1,$... . The following counter example: $f(t) = \exp \{ t \}$
shows that this theorem is wrong for the distributions from
$D^{\prime }({\bf R}_{+})$. S.Bernstein \cite{ber} studied the absolutely
monotonic functions. It is natural to have a try to generalize the
Bernstein result. We shall prove that if for a distribution
$f(t) \in D^{\prime }({\bf R}_{+})$ a distribution
$g(t) = f(- t) \in D^{\prime }({\bf R}_{-})$  is absolutely monotonic then

\begin{equation}
\label{1.4}
f(t) = \int_{0}^{\infty} e^{- ts} d\mu (s),
\end{equation}
where the positive measure $\mu (s)$ has tempered growth. The measure
$\mu (s)$  explicitely depends on the distribution $f(t)$. It is the
sum of two limits of the special distributions constructed from the
distribution $f(t)$. By using the generalized Bernstein theorem it is
possible to obtain the new inversion formula for the Laplace
transformation of the tempered distributions with supports in the
closed positive semiaxis. Our weak spectral condition on the Euclidean
Green's functions requires that some of the limits needed for the
inversion formula exist for any Euclidean Green's function with even
number of variables. We shall prove that the initial
Osterwalder--Schrader axioms \cite{os1}  and the weak spectral condition
are equivalent with the Wightman axioms.

In the next section we study the absolutely monotonic distributions.
The generalization of the Bernstein theorem \cite{ber} is proved. The
new inversion formula for the Laplace transformation of the tempered
distributions with supports in the closed positive semiaxis is obtained.
The third section is devoted to study the exponentially convex tempered
distributions and the tempered distributions satisfying the
Osterwalder--Schrader positivity condition which includes the space
variables. In the fourth section the revised Osterwalder--Schrader
theorem is proved.

\section{Absolutely monotonic distributions}
\setcounter{equation}{0}

\noindent $D({\bf R}_{+})$ denotes the subspace of $D({\bf R})$ of
functions with support in the positive semiaxis
$\overline{{\bf R}}_{+} = [0,\infty )$, given the induced topology.
Similarly $D({\bf R}_{-})$ denotes the subspace of $D({\bf R})$ of
functions with support in the negative semiaxis
$\overline{{\bf R}}_{-} = (-\infty ,0]$, given the induced topology.
If the function $\phi (x) \in D({\bf R}_{-})$ then the function
$\phi (-x) \in D({\bf R}_{+})$.

Since the topology of the space $D({\bf R}_{+})$  is induced by the
topology of the space $D({\bf R})$ there exists a natural number
$N$ for a distribution $f \in D^{\prime }({\bf R}_{+})$  such that
the estimation

\begin{equation}
\label{2.1}
|(f,\phi )| \leq C \sup_{x \in {\bf R}_{+}, 0 \leq k < N}
|\frac{d^{k}\phi }{dx^{k}} |
\end{equation}
holds for every function $\phi (x) \in D({\bf R})$ with support
in the interval $[0,1]$. By using the estimation (\ref{2.1})
and the identity

\begin{equation}
\label{2.2}
\frac{d^{k}}{dx^{k}} (x^{N_{1}}\phi (x)) =  x^{N_{1} - k -1}
\bigl[ \prod_{j = 1}^{k} (N_{1} - j + x\frac{d}{dx} ) \bigr]
(x\phi (x))
\end{equation}
for a natural number $k$ it is easy to show
for any integer $N_{1} \geq N$ that the inequality

\begin{equation}
\label{2.3}
|(x^{N_{1}}f(x),\phi (x))| \leq C_{1}
\sup_{x \in {\bf R}_{+}, 0 \leq k < N}
|(x\frac{d}{dx})^{k} (x\phi (x)) |
\end{equation}
holds for every function $\phi (x) \in D({\bf R})$ with support in
the interval $[0,1]$. We denote by $N(f)$ the minimal integer $N_{1}$
such that the distribution $x^{N_{1}}f(x) \in D^{\prime }({\bf R}_{+})$
satisfies the inequality of type (\ref{2.3}) for every function
$\phi (x) \in D({\bf R})$ with support in the interval $[0,1]$. The
inequality (\ref{2.3}) implies the inequality

\begin{equation}
\label{2.4}
N(f) \geq N(x\frac{df}{dx})
\end{equation}
for any distribution $f(x) \in D^{\prime }({\bf R}_{+})$.

\noindent {\bf Lemma~2.1.} {\it For every distribution}
$f(x) \in D^{\prime }({\bf R}_{+})$,  {\it for any function}
$\phi (x) \in D({\bf R})$ {\it and for every integer} $N \geq N(f)$
{\it the limit}

\begin{equation}
\label{2.5}
\lim_{\alpha \rightarrow + 0} (x^{N}f(x),\theta (x)
\exp \{ -\alpha x^{- 1}\} \phi (x))
\end{equation}
{\it defines the extension} $[x^{N}f](x) \in D^{\prime }({\bf R})$
{\it with support in the positive semiaxis} $\overline{{\bf R}}_{+} $.

\noindent {\it Proof.} Lemma 2.1 follows from the estimation (\ref{2.3})
and \cite[equality (14)]{zin}.

The distribution $f(x) \in D^{\prime }({\bf R}_{-})$  is said
to be absolutely monotonic if for all natural numbers $m = 0,1$,...
the distribution $\frac{d^{m}f}{dx^{m}} (x)$ is positive.

If a function $\phi (x) \in D({\bf R})$ then for sufficiently large
positive $t$ the function $\phi (x + t) \in D^{\prime }({\bf R}_{-})$.

\noindent {\bf Lemma~2.2.} {\it Let the distribution}
$f(x) \in D^{\prime }({\bf R}_{-})$ {\it be absolutely monotonic then for
any function} $\phi (x) \in D({\bf R})$

\begin{equation}
\label{2.7}
\lim_{t \rightarrow - \infty} (f(x),\phi (x - t)) =
L_{0}^{- 1}[f] \int_{- \infty}^{\infty} \phi (x) dx
\end{equation}

\begin{equation}
\label{2.8}
\lim_{t \rightarrow - \infty} t^{k}(\frac{d^{k}f}{dx^{k}}(x),\phi (x - t))
= 0 ,k = 1,2,...
\end{equation}
{\it If the distribution} $f(x) \in D^{\prime }({\bf R}_{-})$ {\it is
absolutely monotonic the distribution}
$(- x)^{- 1}f(x) \in D^{\prime }({\bf R}_{-})$ {\it is also absolutely
monotonic, and the constant} $L_{0}^{- 1}[(- x)^{- 1}f(x)] = 0$.

\noindent {\it Proof.} Let the function $\phi (x) \in D({\bf R})$
be positive and its support be in the interval $[a,b]$. The function
$f(t;\phi ) \equiv (f(x),\phi (x - t))$ is defined on the semiaxis
$(- \infty, - a]$. It is infinitely differentiable. Since the
distribution $f(x)$ is absolutely monotonic the positivity of the
function $\phi (x)$ implies

\begin{equation}
\label{2.9}
\frac{d^{n}}{dt^{n}} f(t;\phi ) =
(\frac{d^{n}f}{dx^{n}}(x),\phi (x - t)) \geq 0, n = 0,1,...
\end{equation}
Hence for every $n = 1,2$,... we get

\begin{eqnarray}
\label{2.10}
& & \frac{d^{n}}{dt^{n}} f(t;\phi ) \leq 2|t|^{- 1} \int_{t}^{t/2}
\frac{d^{n}}{dy^{n}} f(y;\phi ) dy = \\
& & \leq 2|t|^{- 1} \left[ \frac{d^{n - 1}}{dy^{n - 1}} f(y;\phi )
\mid_{y = t/2} - \frac{d^{n - 1}}{dt^{n - 1}} f(t;\phi )\right] ,
\nonumber
\end{eqnarray}
where $t < 0, - a$. Due to the inequlities (\ref{2.9}) the function
$f(t;\phi )$ is positive and non--decreasing on the semiaxis
$(- \infty , - a]$. Therefore the limit (\ref{2.7}) exists. Then it
follows from the inequality (\ref{2.10}) for $n = 1$ that the limit
(\ref{2.8}) equals zero for $k = 1$. By using the induction and the
inequality (\ref{2.10}) it is easy to prove the equalities (\ref{2.8})
for $k = 1,2$,... and for any positive function $\phi (x) \in D({\bf R})$.

Let a function $\phi (x) \in D({\bf R})$ and the number
$M = \sup_{x \in {\bf R}} |\phi (x)|$. Let a positive function
$h(x) \in D({\bf R})$ be equal to one on the support of the function
$\phi (x)$. The function $\phi (x)$ is the difference of the positive
functions $1/2 (Mh(x) \pm \phi (x))$. This decomposition implies the
equalities (\ref{2.8}) for the function $\phi (x)$ since the equalities
(\ref{2.8}) are valid for the positive functions from $D({\bf R})$.

Let the integral of a positive function $h(x) \in D({\bf R})$ be equal to one.
Then any function $\phi (x) \in D({\bf R})$ may be rewritten as

\begin{equation}
\label{2.11}
\phi (x) = h(x)\int_{- \infty}^{\infty} \phi (y) dy  +
\frac{d\psi }{dx} (x),
\end{equation}
where $\psi (x) \in D({\bf R})$. The limit (\ref{2.7})  exists for any
positive function from $D({\bf R})$. Hence the decomposition (\ref{2.11})
and the equality (\ref{2.8}) for $k = 1$  imply the equality (\ref{2.7}).

If the distribution $f(x) \in D^{\prime }({\bf R}_{-})$ is absolutely
monotonic the distribution $(- x)^{- 1}f(x) \in D^{\prime }({\bf R}_{-})$
is also absolutely monotonic. It follows from the relations (\ref{2.7})
for the distributions $f(x)$ and $(- x)^{- 1}f(x)$ that

\begin{eqnarray}
\label{2.6}
& & \lim_{t \rightarrow -\infty } t((- x)^{- 1}f(x),\phi (x - t)) = \\
& & - L_{0}^{- 1}[(- x)^{- 1}f(x)] \int_{- \infty }^{\infty } x\phi (x) dx -
L_{0}^{- 1}[f] \int_{- \infty }^{\infty } \phi (x) dx \nonumber
\end{eqnarray}
Since the limit (\ref{2.7}) for the distribution $(- x)^{- 1}f(x)$
exists the limit (\ref{2.6}) may exist when the constant
$L_{0}^{- 1}[(- x)^{- 1}f(x)] = 0$.

For a function $\phi (x) \in D({\bf R})$ and $k = 1,2$,... we introduce
the function

\begin{equation}
\label{2.12}
\phi^{(- k)} (x) = - ((k - 1)!)^{- 1} \int_{x}^{\infty } (x - y)^{k - 1}
\phi (y) dy
\end{equation}
and for $k = 0$ we define $\phi^{(0)} (x) = \phi (x)$. The infinitely
differentiable function (\ref{2.12}) equals zero on the positive
semiaxis for $\phi (x) \in D({\bf R}_{-})$. Our notation is reasonable
since $\frac{d^{l}}{dx^{l}} \phi^{(- k)} (x) = \phi^{(l - k)} (x)$  for
$l \leq k$.

Let a positive function $h(x) \in D({\bf R})$ have the integral equal
one and its support be in the positive semiaxis. Let us construct
the infinitely differentiable function with finite support for every $T < 0$

\begin{equation}
\label{2.13}
h_{T} (x) = \int_{x}^{x - T} h(y)dy.
\end{equation}

\noindent {\bf Lemma~2.3.} {\it Let a distribution}
$f(x) \in D^{\prime }({\bf R}_{-})$ {\it be absolutely monotonic. Then for
any function} $\phi (x) \in D({\bf R}_{-})$ {\it and for any integer}
$k = 1,2$,...

\begin{equation}
\label{2.14}
\lim_{T \rightarrow - \infty } (- 1)^{k}
(\frac{d^{k}f}{dx^{k}} (x),h_{T}(x) \phi^{(- k)} (x)) =
(f(x),\phi (x)) - L_{0}^{- 1}[f] \int_{- \infty}^{\infty} \phi (x) dx ,
\end{equation}
{\it where the constant} $L_{0}^{- 1}[f]$ {\it is given by the equality}
(\ref{2.7}).

\noindent {\it Proof.} The definitions (\ref{2.12}) and (\ref{2.13})
imply for any integer $k = 1,2$,... the following relation

\begin{eqnarray}
\label{2.15}
& & (\frac{d^{k}f}{dx^{k}} (x),h_{T}(x) \phi^{(- k)} (x)) = -
(\frac{d^{k - 1}f}{dx^{k - 1}} (x),h_{T}(x) \phi^{(1 - k)} (x)) \nonumber \\
& & - (\frac{d^{k - 1}f}{dx^{k - 1}} (x),(h(x - T) - h(x))\phi^{(- k)} (x)).
\end{eqnarray}

Due to the definitions the supports of the functions $h(x)$ and
$\phi^{(- k)} (x)$ don't intersect. Hence the function
$h(x)\phi^{(- k)} (x) = 0$. By using the equality (\ref{2.15}) $k$
times we get

\begin{eqnarray}
\label{2.16}
& & (- 1)^{k}(\frac{d^{k}f}{dx^{k}} (x),h_{T}(x) \phi^{(- k)} (x)) =
(f(x),h_{T}(x) \phi (x)) + \nonumber \\
& & \sum_{p = 0}^{k - 1}
(- 1)^{p}(\frac{d^{p}f}{dx^{p}} (x),h(x - T) \phi^{(- p - 1)} (x)).
\end{eqnarray}
Since the supports of the functions $h(x)$ and $\phi (x)$ are finite
we obtain for the sufficiently large modulus of the negative number $T$

\begin{equation}
\label{2.17}
h(x - T)\int_{- \infty }^{x} (x - y)^{p - 1}\phi (y) dy = 0,
\end{equation}
where the integer $p = 1,2$,... . It follows from the relations
(\ref{2.7}), (\ref{2.8}), (\ref{2.12}) and (\ref{2.17}) that

\begin{equation}
\label{2.18}
\lim_{T \rightarrow - \infty } \sum_{p = 0}^{k - 1}
(- 1)^{p}(\frac{d^{p}f}{dx^{p}} (x),h(x - T) \phi^{(- p - 1)} (x)) =
- L_{0}^{- 1}[f] \int_{- \infty }^{\infty } \phi (x) dx ,
\end{equation}
where the constant $L_{0}^{- 1}[f]$ is given by the equality (\ref{2.7}).

The definition (\ref{2.13}) implies that the function $h_{T}(x)$ is
equal to one on the support of the function $\phi (x) \in D({\bf R}_{-})$
for the sufficiently large modulus of the negative number $T$, as the
integral of the function $h(x) \in D({\bf R}_{+})$ is equal to one.
Now the equality (\ref{2.14}) is the consequence of the equalities
(\ref{2.16}) and (\ref{2.18}).

For a distribution $f(x) \in D^{\prime }({\bf R}_{-})$ we define a
functional on the space $S({\bf R})$ by the following relation

\begin{eqnarray}
\label{2.19}
& & (L_{c}^{- 1}[f](-x; n,T),\phi (x)) =
(f(x),L_{c}^{- 1}[\phi ](-x; n,T)) = \nonumber \\
& & (n!)^{- 1} ((- x)^{n}\frac{d^{n + 1}f}{dx^{n + 1}} (x), \theta (- x)
h_{T}(x)\phi (- nx^{- 1})),
\end{eqnarray}
where $n$ is a positive integer and the function $h_{T}(x)$ is given by
the equality (\ref{2.13}). It is easy to show that the tempered
distribution $L_{c}^{- 1}[f](-x; n,T) \in S^{\prime }({\bf R})$ is
positive and its support is in the positive semiaxis.

\noindent {\bf Proposition~2.4.} {\it Let a distribution}
$f(x) \in D^{\prime }({\bf R}_{-})$ {\it be absolutely monotonic. Then in
the topological space} $S^{\prime }({\bf R})$ {\it there exists the limit}

\begin{equation}
\label{2.20}
\lim_{n \rightarrow \infty } \lim_{T \rightarrow - \infty }
L_{c}^{- 1}[f](-x; n,T) = L_{c}^{- 1}[f](-x)
\end{equation}
{\it The tempered distribution} $L_{c}^{- 1}[f](-x) \in S^{\prime }({\bf R})$
{\it is positive and its support is in the positive semiaxis.}

\noindent {\it Proof.} Let us multiply and divide the function $\phi (x)$
in the right-hand side of the equalities (\ref{2.19}) by the same
polynomial $(1 + x)^{N}$

\begin{eqnarray}
\label{2.21}
& & (L_{c}^{- 1}[f](-x; n,T),\phi (x)) =
(n!)^{- 1} \int dx (- x)^{n + N(f) + 1}\frac{d^{n + 1}f}{dx^{n + 1}} (x)
\times \nonumber \\
& & (- x)^{N - N(f) - 1}\theta (- x)h_{T}(x)
\bigl( \frac{1 - nx^{- 1}}{n - x} \bigr)^{N} \phi (- nx^{- 1}),
\end{eqnarray}
where $N(f)$ is the minimal integer $N_{1}$ such that the inequality
(\ref{2.3}) is satisfied. The relation (\ref{2.2}) for $N_{1} = 1$
implies

\begin{equation}
\label{2.22}
(- x)^{n + N(f) + 1}\frac{d^{n + 1}f}{dx^{n + 1}} (x) =
(- 1)^{n + N(f) + 1}(x)^{N(f) + 1}
\bigl( \prod_{j = 1}^{n} (x\frac{d}{dx} + 1 - j) \bigr)
\frac{df}{dx} (x) .
\end{equation}
Due to the inequality (\ref{2.4}) and Lemma 2.1 the positive
distribution (\ref{2.22}) from $D^{\prime }({\bf R}_{-})$ is
extended to the positive distribution from $D^{\prime }({\bf R})$
with support in the negative semiaxis. This extension is defined by
the limit analogous to the limit (\ref{2.5}).

For the sufficiently large positive integer $n$  the function
$(- x)^{N - N(f) - 1}h_{T}(x)(n - x)^{- N}$ is infinitely differentiable
for $N > N(f)$. It is positive on the negative semiaxis. Now the
positivity of the extension of the distribution (\ref{2.22}) implies
the following estimation of the integral (\ref{2.21})

\begin{equation}
\label{2.23}
|(L_{c}^{- 1}[f](-x; n,T),\phi (x))| \leq C_{n,T}(N)
\sup_{x \geq 0} (1 + x)^{N} |\phi (x)| .
\end{equation}
Here the constant

\begin{eqnarray}
\label{2.24}
& & C_{n,T}(N) =
(n!)^{- 1} \lim_{\alpha \rightarrow + 0} \int dx (- x)^{n + N(f) + 1}
\frac{d^{n + 1}f}{dx^{n + 1}} (x) \times \nonumber \\
& & \theta (- x)\exp \{ \alpha nx^{- 1}\}
(- x)^{N - N(f) - 1}h_{T}(x)(n - x)^{- N} = \nonumber \\
& & \lim_{\alpha \rightarrow + 0} (L_{c}^{- 1}[f](-x; n,T),
\exp \{ - \alpha x\} (1 + x)^{- N})
\end{eqnarray}
and the integer $N > N(f)$. We denote by $|\bullet |_{N}$ the norm in
the right-hand side of the inequality (\ref{2.23}). We define $H_{N}$
to be the Bahach space completion of the space $S({\bf R})$. Due to the
inequality (\ref{2.23}) for $N = N(f) + 1$ the tempered distribution
$L_{c}^{- 1}[f](-x; n,T) \in S^{\prime }({\bf R})$ is continued to the
linear continuous functional $L_{c}^{- 1}[f](-x; n,T)^{c}$ on the Banach
space $H_{N(f) + 1}$. Now the relation (\ref{2.24}) may be rewritten
as $C_{n,T}(N) = (L_{c}^{- 1}[f](-x; n,T)^{c},(1 + x)^{- N})$,
where $N > N(f)$.

Let us assume that for any integer $N > N(f)$ there exists the limit

\begin{equation}
\label{2.25}
\lim_{n \rightarrow \infty } \lim_{T \rightarrow - \infty }
C_{n,T}(N)
\end{equation}
This assumption and the inequality (\ref{2.23}) for $N = N(f) + 1$
imply that the linear continuous functionals $L_{c}^{- 1}[f](-x; n,T)^{c}$
on the Banach space $H_{N(f) + 1}$ are uniformly bounded. In view of the
relation (\ref{2.24}) the existence of the limit (\ref{2.25}) is
equivalent to the covergence of sequence $\{ L_{c}^{- 1}[f](-x; n,T)^{c}\} $
on every function $(1 + x)^{- N} \in H_{N(f) + 1}$, where $N > N(f)$.
If the set of these functions is dense in the Banach space $H_{N(f) + 1}$
then by the Banach--Steinhaus theorem \cite[Section 3.7]{vla} the
sequence of linear continuous functionals $L_{c}^{- 1}[f](-x; n,T)^{c}$
on the Banach space $H_{N(f) + 1}$ weakly converges to the linear
continuous functional on $H_{N(f) + 1}$. Therefore the sequence of
tempered distributions $L_{c}^{- 1}[f](-x; n,T) \in S^{\prime }({\bf R})$
weakly converges to the tempered distribution from $S^{\prime }({\bf R})$.
Hence due to \cite[Section 3.7]{vla} the sequence of tempered
distributions $L_{c}^{- 1}[f](-x; n,T)$ coverges in topology of the
space $S^{\prime }({\bf R})$. Now it is easy to prove that the tempered
distribution $L_{c}^{- 1}[f](-x) \equiv $
$L_{c}^{- 1}[f](-x; \infty , - \infty )$$\in S^{\prime }({\bf R})$ is
positive and its support is in the positive semiaxis.

Let us prove at first that the set of the functions
$(1 + x)^{- N(f) - k - 1}$, $k = 0,1$,... , is dense in the Banach
space $H_{N(f) + 1}$. For any function $\phi (x) \in S({\bf R})$ the
function $\hat{\phi } (t) = \phi (\tan^{2} t) (\cos t)^{- 2(N(f) + 1)}$
is continuous on the interval $(- \pi /2 ,\pi /2)$. Due to a function
$\phi (x) \in S({\bf R})$ the function $\hat{\phi } (t)$ may be
continued on the closed interval $[- \pi /2 ,\pi /2]$ by setting
$\hat{\phi } (- \pi /2) = \hat{\phi } (\pi /2) = 0$. Hence the
Weierstrass theorem implies that the periodic continuous function
$\hat{\phi } (t)$ on the closed interval $[- \pi /2 ,\pi /2]$ is
approximated by the trigonometric polynomial $\sum b_{m} \exp \{ 2mit\}$.
Since the function $\hat{\phi } (t)$ is even it is approximated by
the trigonometric polynomial $\sum b_{m} \cos 2mt$. The function
$\cos 2mt$ is the polynomial of the variable $\cos^{2} t$.
Therefore for every positive $\epsilon $  there exists the polynomial
$\sum a_{m}\cos^{2m} t$  such that the modulus of the function
$\hat{\phi } (t) - \sum a_{m}\cos^{2m} t$ on the closed interval
$[- \pi /2 ,\pi /2]$ is less than $\epsilon $. Due to the relation
$\cos^{2m} t = (1 + \tan^{2} t)^{- m}$ it implies that

\begin{equation}
\label{2.26}
\sup_{x \geq 0} (1 + x)^{N(f) + 1} |\phi (x) -
\sum a_{m}(1 + x)^{- N(f) - m - 1}| < \epsilon
\end{equation}
Thus the set of the functions $(1 + x)^{- N(f) - k - 1}$, $k = 0,1$,... ,
is dense in the Banach space $H_{N(f) + 1}$, as the space $S({\bf R})$
is dense.

At last let us prove the existence of the limit (\ref{2.25}) for
every integer $N > N(f)$. If a function $h(x) \in D({\bf R}_{+})$
then for sufficiently large modulus of the negative number $T$
a function $h(x - T) \in D({\bf R}_{-})$. By using the definition
(\ref{2.13}), the identity (\ref{2.2}) and the estimation
(\ref{2.3}) we can rewrite the expression (\ref{2.24}) for
sufficiently large modulus of the negative number $T$ in the
following form

\begin{equation}
\label{2.27}
C_{n,T}(N) = B_{n,T}(N) -
(n!)^{- 1} \sum_{k = 0}^{n - 1} (- 1)^{n - k}
(\frac{d^{n - k}f}{dx^{n - k}} (x),h(x - T)
\frac{d^{k}}{dx^{k}} (x^{n + N}(x - n)^{- N})),
\end{equation}
where the constant

\begin{equation}
\label{2.28}
B_{n,T}(N) = \lim_{\alpha \rightarrow + 0} (\frac{df}{dx} (x),
\theta (- x) \exp \{ \alpha x^{- 1}\} h_{T} (x) \chi_{n,N} (- x^{- 1}))
\end{equation}
and the function

\begin{equation}
\label{2.29}
\chi_{n,N} (x) = (n!)^{- 1}\frac{d^{n}}{dy^{n}}
(y^{n + N}(y - n)^{- N}) \mid_{y = - x^{- 1}} .
\end{equation}
It follows from the identity (\ref{2.2}) that

\begin{equation}
\label{2.30}
x^{k - n}\frac{d^{n}}{dx^{n}} (x^{n + N}(x - n)^{- N}) =
\bigl[ \prod_{j = 1}^{k} (n + 1 - j - y\frac{d}{dy} ) \bigr]
(1 + y)^{- N} \mid_{y = - nx^{- 1}} .
\end{equation}
Now it is easy to show that the expression (\ref{2.30}) is bounded
on the closed negative semiaxis. Hence the absolutely monotonicity of
the distribution $f(x) \in D^{\prime }({\bf R}_{-})$ and the relations
(\ref{2.8}), (\ref{2.27}) imply that

\begin{equation}
\label{2.31}
\lim_{T \rightarrow - \infty } C_{n,T}(N) =
\lim_{T \rightarrow - \infty } B_{n,T}(N) .
\end{equation}
Let us prove that the numbers $B_{n,T}(N)$ for $N > N(f)$ form the
Cauchy sequence when $n \rightarrow \infty $, $T \rightarrow - \infty $.
In view of the equality (\ref{2.31}) it implies the existence of
the limit (\ref{2.25}). If $T_{2} \leq T_{1}$, then for the
sufficiently large modulus of the negative number $T_{1}$ the
positive function $h_{T_{2}} (x) - h_{T_{1}} (x) \in D({\bf R}_{-})$
and by the positivity of the distribution
$\frac{df}{dx} (x) \in D^{\prime }({\bf R}_{-})$ the relation (\ref{2.28})
implies the following estimation

\begin{equation}
\label{2.32}
|B_{n,T_{2}}(N) - B_{n,T_{1}}(N)| \leq
(f(x),h(x - T_{1}) - h(x - T_{2}))
\sup_{x \geq 0} |\chi_{n,N} (x)|.
\end{equation}
In virtue of relation (\ref{2.7}) the first multiplier in the
right--hand side of the inequality (\ref{2.32}) converges to zero
when $T_{1}, T_{2} \rightarrow - \infty $. Due to the positivity
of the distribution $\frac{df}{dx} (x) \in D^{\prime }({\bf R}_{-})$
we find from the relation (\ref{2.28})

\begin{eqnarray}
\label{2.33}
& & |B_{n_{2},T}(N) - B_{n_{1},T}(N)| \leq \\
& & \lim_{\alpha \rightarrow + 0} (\frac{df}{dx} (x),
\theta (- x) \exp \{ \alpha x^{- 1}\} h_{T} (x)
(1 - x^{- 1})^{- N(f) - 1}) \times \nonumber \\
&  & \sup_{x \geq 0} (1 + x)^{N(f) + 1}
|\chi_{n_{2},N} (x) - \chi_{n_{1},N} (x)|.
\nonumber
\end{eqnarray}
We denote the first multiplier in the right--hand side of the
inequality (\ref{2.33}) by $A(T)$. The numbers $A(T)$ form the
Cauchy sequence when $T \rightarrow - \infty $ and consequently the
numbers $A(T)$ are uniformly bounded on the negative semiaxis.
The proof of this is exactly analogous to that of the inequality
(\ref{2.32}). If we prove that the functions (\ref{2.29})
converge in norm $|\bullet |_{N(f) + 1}$ to the function

\begin{equation}
\label{2.34}
\chi_{\infty ,N} (x) = ((N - 1)!)^{- 1} \int_{0}^{x^{- 1}}
t^{N - 1} e^{- t}dt
\end{equation}
when $n \rightarrow \infty $, the inequalities (\ref{2.32}) and
(\ref{2.33}) provide us that the numbers $B_{n,T}(N)$ form the
Cauchy sequence when $n \rightarrow \infty $, $T \rightarrow - \infty $.

By the definition (\ref{2.29}) we get

\begin{eqnarray}
\label{2.35}
& & \chi_{n,N} (x) = \frac{(n + N)!}{n!(N - 1)!} \sum_{k = 0}^{n}
(- 1)^{n - k} \frac{n!}{k!(n - k)!} \frac{(1 + nx)^{- n - N + k}}{n + N - k}
= \nonumber \\
& & \frac{(n + N)!}{n^{N}n!(N - 1)!} \int_{0}^{(n^{-1} + x)^{-1}}
t^{N - 1}(1 - tn^{-1})^{n} dt.
\end{eqnarray}
The inequality $t^{N - 1}e^{- t} \leq x^{- N - 1}t^{- 2}$ holds on the
interval $[(n^{-1} + x)^{-1},x^{-1}]$, where $x,n > 0$. It implies
the following estimation

\begin{equation}
\label{2.36}
\sup_{1 \leq x < \infty } (1 + x)^{N(f) + 1}
\int_{(n^{-1} + x)^{-1}}^{x^{-1}}
t^{N - 1} e^{- t} dt \leq n^{-1} 2^{N(f) + 1} .
\end{equation}
Here we use the inequality
$(1 + x)^{N(f) + 1}x^{- N - 1} \leq 2^{N(f) + 1}$ valid on the
semiaxis $[1,\infty )$ for $N \geq N(f)$. The maximal value of the
function $t^{N + 1} e^{- t}$ on the positive semiaxis is equal to
$((N + 1)e^{-1})^{N + 1}$. Hence by using the inequality
$(1 + x)^{N(f) + 1} \leq 2^{N(f) + 1}$ for $0 \leq x \leq 1$ we
obtain

\begin{equation}
\label{2.37}
\sup_{0 \leq x \leq 1} (1 + x)^{N(f) + 1}
\int_{(n^{-1} + x)^{-1}}^{x^{-1}}
t^{N - 1} e^{- t} dt \leq n^{-1} 2^{N(f) + 1} ((N + 1)e^{-1})^{N + 1}.
\end{equation}
For any positive number $x$ and for a natural number $n > 0$
the number $(n^{-1} + x)^{-1} \leq n$. Then the equalities
(\ref{2.34}), (\ref{2.35}) and the estimation (\ref{2.37})
imply that

\begin{eqnarray}
\label{2.38}
& & \sup_{0 \leq x \leq 1} (1 + x)^{N(f) + 1}
|\chi_{\infty ,N} (x) - \frac{n^{N}n!}{(n + N)!} \chi_{n,N} (x)| \leq
\nonumber \\
& & 2^{N(f) + 1} ((N - 1)!)^{-1} \bigl[ n^{-1} ((N + 1)e^{-1})^{N + 1} + \\
& & \int_{0}^{\infty } (1 + s^{2})^{-1} ds \sup_{0 \leq t < \infty }
(1 + t^{2}) t^{N - 1} |e^{- t} - (1 - tn^{-1})_{+}^{n}| \bigr] ,
\nonumber
\end{eqnarray}
where the function $x_{+}^{n}$ is equal to $x^{n}$ for $x \geq 0$
and it is equal to zero, otherwise. It follows from the equalities
(\ref{2.34}), (\ref{2.35}) and the estimation (\ref{2.36})
for $N > N(f)$ that

\begin{eqnarray}
\label{2.39}
& & \sup_{1 \leq x < \infty } (1 + x)^{N(f) + 1}
|\chi_{\infty ,N} (x) - \frac{n^{N}n!}{(n + N)!} \chi_{n,N} (x)| \leq  \\
& & 2^{N(f) + 1} ((N - 1)!)^{-1} \left[ n^{-1}  +
N^{-1} \sup_{0 \leq t < \infty } |e^{- t} - (1 - tn^{-1})_{+}^{n}| \right] .
\nonumber
\end{eqnarray}
For a natural number $n > 1$ the function $x_{+}^{n}$ is differentiable
everywhere. The maximal value of the function
$\exp \{ - t\} - (1 - tn^{-1})_{+}^{n}$  on the positive semiaxis
is at the point $a_{0}$ satisfying the equation
$\exp \{ - a_{0}\} = (1 - a_{0}n^{-1})_{+}^{n - 1}$. This equation
implies

\begin{equation}
\label{2.40}
\sup_{0 \leq t < \infty } |e^{- t} - (1 - tn^{-1})_{+}^{n}| =
|e^{- a_{0}} - (1 - a_{0}n^{-1})_{+}^{n}| =
n^{-1}a_{0}e^{- a_{0}} \leq (ne)^{-1}.
\end{equation}
For the natural numbers $k$ and $n > 1$ the maximal value of the
function $t^{k}(e^{- t} - (1 - tn^{-1})_{+}^{n})$ on the positive
semiaxis is at the point $a_{k}$ satisfying the following equation

\begin{equation}
\label{2.41}
ka_{k}^{k - 1}(e^{- a_{k}} - (1 - a_{k}n^{-1})_{+}^{n}) =
a_{k}^{k}(e^{- a_{k}} - (1 - a_{k}n^{-1})_{+}^{n - 1}).
\end{equation}
If $a_{k} \leq k + 1$ the inequality (\ref{2.40}) implies the
estimation

\begin{equation}
\label{2.42}
\sup_{0 \leq t < \infty } t^{k}|e^{- t} - (1 - tn^{-1})_{+}^{n}| \leq
(k + 1)^{k}(ne)^{-1}.
\end{equation}
If $a_{k} \geq k + 1$ by using the equation (\ref{2.41}) we get

\begin{eqnarray}
\label{2.43}
& & \sup_{0 \leq t < \infty } t^{k}|e^{- t} - (1 - tn^{-1})_{+}^{n}| =
n^{-1}((1 + n^{-1}k)a_{k} - k)^{-1}a_{k}^{k + 2}e^{- a_{k}} < \nonumber \\
& &n^{-1}a_{k}^{k + 2}e^{- a_{k}} \leq
n^{-1} ((k + 2)e^{-1})^{k + 2} .
\end{eqnarray}
In order to get the first inequality (\ref{2.43}) we use the
inequality $(1 + n^{-1}k)a_{k} - k > 1$ valid for $a_{k} \geq k + 1$.
It follows from the estimations (\ref{2.38}), (\ref{2.39}),
(\ref{2.40}), (\ref{2.42}) and (\ref{2.43}) that the functions
(\ref{2.35}) converge in norm $|\bullet |_{N(f) + 1}$ to the function
(\ref{2.34}). Therefore Proposition 2.4 is proved.

By the definitions (\ref{2.19}) and (\ref{2.20}) the tempered
distribution $L_{c}^{- 1}[f](-x) \in S^{\prime }({\bf R})$ is
positive and its support is in the positive semiaxis. Now
Theorem 2 from \cite[Chapter 2, Section 2.2]{gvi} implies that
the positive tempered distribution is given by the positive measure.
This positive measure has tempered growth and its support is in the
positive semiaxis.

\noindent {\bf Theorem~2.5.} {\it For any absolutely monotonic
distribution} $f(x) \in D^{\prime }({\bf R}_{-})$
{\it the following representation}

\begin{equation}
\label{2.44}
(f(x),\phi (x)) = L_{0}^{- 1}[f] \int_{- \infty }^{\infty } \phi (x) dx +
\int_{- \infty }^{\infty } L_{c}^{- 1}[f](-p) dp
\int_{- \infty }^{\infty } e^{px} \phi (x) dx
\end{equation}
{\it is valid for any function} $\phi (x) \in D({\bf R}_{-})$.
{\it Here the positive number} $L_{0}^{- 1}[f]$ {\it is given by
the relation} (\ref{2.7}), {\it the positive measure}
$L_{c}^{- 1}[f](-p)$ {\it with tempered growth and support in the positive
semiaxis is defined by the relations} (\ref{2.19}), (\ref{2.20}).

\noindent {\it Proof.} Let a distribution
$f(x) \in D^{\prime }({\bf R}_{-})$ be absolutely monotonic. Then in
view of Lemma 2.3 the relation (\ref{2.14}) holds for any function
$\phi (x) \in D({\bf R}_{-})$ and for any integer $k = n + 1$, where
$n$ is a natural number. Note that

\begin{eqnarray}
\label{2.45}
& & (- 1)^{n + 1}(\frac{d^{n + 1}f}{dx^{n + 1}} (x),
h_{T}(x) \phi^{(- n - 1)} (x)) = \\
& & (n!)^{-1} ((- x)^{n}\frac{d^{n + 1}f}{dx^{n + 1}} (x),
h_{T}(x) L_{n}[\phi ](- nx^{-1})) ,\nonumber
\end{eqnarray}
where the function

\begin{equation}
\label{2.46}
L_{n}[\phi ](x) = \int_{- nx^{-1}}^{0} (1 + n^{-1}xy)^{n}\phi (y) dy .
\end{equation}
We take into account that the function $\phi (x)$ has the support in the
negative semiaxis.

Due to the inequality (\ref{2.23}) for $N = N(f) + 1$ the tempered
distribution $L_{c}^{- 1}[f](-x; n,T)$, defined by the equality
(\ref{2.19}), is continued to the linear continuous functional
$L_{c}^{- 1}[f](-x; n,$$T$$)^{c}$ on the Banach $H_{N(f) + 1}$.
Let us prove that the function (\ref{2.46}) converge as
$n \rightarrow \infty $ in norm $|\bullet |_{N(f) + 1}$ to the
Laplace transform of the function $\phi (- x)$. It is straightforward
to show that

\begin{eqnarray}
\label{2.47}
& & |\int_{- \infty }^{0} e^{xy}\phi (y) dy - L_{n}[\phi ](x)|_{N(f) + 1}
\leq \\
& & \int_{- \infty }^{0} |\phi (y)| dy \sup_{0 \leq x < \infty }
(1 + x)^{N(f) + 1}|e^{xy} - (1 + n^{-1}xy)_{+}^{n}|.\nonumber
\end{eqnarray}
The right--hand side of the inequality (\ref{2.47}) is majorized
by the sum

\begin{equation}
\label{2.48}
\sum_{k = 0}^{N(f) + 1} \frac{(N(f) + 1)!}{k!(N(f) + 1 - k)!}
\int_{- \infty }^{0} |y|^{- k}|\phi (y)| dy \sup_{0 \leq t < \infty }
t^{k}|e^{t} - (1 + n^{-1}t)_{+}^{n}|.
\end{equation}
Due to the estimations (\ref{2.42}), (\ref{2.43}) the sum (\ref{2.48})
converges to zero as $n \rightarrow \infty $. Thus we have proved
that the right--hand side of the equality (\ref{2.45}) converges to

\begin{equation}
\label{2.49}
(L_{c}^{- 1}[f](-x),\int_{- \infty }^{0} e^{xy}\phi (y) dy).
\end{equation}
Now the relation (\ref{2.14}) implies the equality (\ref{2.44}).

For $O$ an open set in ${\bf R}^{n}$, $S(O)$ denotes the subspace
of $S({\bf R}^{n})$ of functions with support in the closure
$\overline{O} $, given the induced topology. For example $S({\bf R}_{-})$
denotes the subspace of $S({\bf R})$ of functions with support in
the negative semiaxis, given the induced topology. Similarly
$S({\bf R}_{+})$ denotes the subspace of $S({\bf R})$ of functions
with support in the positive semiaxis, given the induced topology.
It follows from Theorem 2.5 that every absolutely monotonic
distribution $f(x) \in D^{\prime }({\bf R}_{-})$ may be extended
to the tempered distribution from $S^{\prime }({\bf R}_{-})$.

We remind that a tempered distribution $g(t) \in S^{\prime }({\bf R}_{-})$
is called absolutely monotonic if it satisfies the conditions
(\ref{1.3}) for all $m = 0,1$,... .

\noindent {  \bf Corollary~2.6.} {\it The tempered distribution}
$f(x) \in S^{\prime }({\bf R}_{+})$ {\it is the Laplace transform
of a tempered distribution with support in the positive semiaxis if
and only if there exists the natural number} $k$ {\it such that}

\begin{equation}
\label{2.50}
(- x)^{- k} f(- x) = g_{1}(x) - g_{2}(x),
\end{equation}
{\it where the tempered distribuions}
$g_{j}(x) \in S^{\prime }({\bf R}_{-})$, $j = 1,2$,
{\it are absolutely monotonic.}

\noindent {\it Proof.} Due to Theorem from \cite[Section 3.8]{vla}
a tempered distribution with support in the positive semiaxis may
be written as

\begin{equation}
\label{2.51}
g(x) = \sum_{m = 0}^{k - 1} \frac{d^{m}}{dx^{m}} \mu_{m} (x),
\end{equation}
where $\mu_{m} (x)$ are the measures with tempered growth and with
supports in the positive semiaxis.

It is easy to verify that $(m!)^{-1}(\frac{d}{dx} )^{m + 1}x_{+}^{m} =$
$ \delta (x)$. Hence the relation (\ref{2.51}) implies

\begin{equation}
\label{2.52}
g(x) =  \frac{d^{k}}{dx^{k}}
[ \sum_{m = 0}^{k - 1} ((k - m - 1)!)^{-1}
x_{+}^{k - m - 1} \ast \mu_{m} (x)] ,
\end{equation}
where $\ast $ denotes the convolution of two tempered distributions
with supports in the positive semiaxis. If we represent the measure
in the right--hand side of the equality (\ref{2.52}) as the
difference of two positive measures with tempered growth and with
supports in the positive semiaxis we get

\begin{equation}
\label{2.53}
g(x) =  \frac{d^{k}}{dx^{k}} [\nu_{1} (x) - \nu_{2} (x)] .
\end{equation}
Taking the Laplace transform of the equality (\ref{2.53}) and
dividing it by $x^{k}$ we obtain the equality (\ref{2.50}), where
$x$ is replaced by $- x$.

It is straightforward to show that the equality (\ref{2.50}) and
Theorem 2.5 imply that the tempered distribution $f(x)$ is the Laplace
transform of a tempered distribution with support in the positive
semiaxis.

Let $x$ denote a point in ${\bf R}^{4}$ with coordinates
$(x^{0},x^{1},x^{2},x^{3}) = (x^{0},{\bf x})$. A point in ${\bf R}^{4n}$
will be written as $\underline{x} = (x_{1},...,x_{n})$,
$x_{i} \in {\bf R}^{4}$. We will use the following open set
${\bf R}_{+}^{4n} = $
$\{ \underline{x} \in {\bf R}^{4n}| x_{j}^{0} > 0, j = 1,...,n \} $.

\noindent {\bf Theorem~2.7.} {\it Let the tempered distribution}
$f(x) \in S^{\prime }({\bf R}_{+}^{4n})$ {\it be the Laplace transform}
{\it with respect to the time variables of a tempered distribution with
support in the closure} $\overline{{\bf R}}_{+}^{4n} $. {\it Then there
is the natural number} $K$ {\it such that for any integers} $k > K$,
$1 \leq j_{1} < \cdots < j_{l} \leq n$, $1 \leq l \leq n$ {\it and for all}
{\it test functions} $\phi_{j} (x) \in S({\bf R}^{4} )$,
$j \in \{ j_{1},...j_{l} \} $; $\phi_{i} (x) \in S({\bf R}_{+}^{4} )$,
$1 \leq i \leq n$, $i \neq j_{1},...j_{l} $, {\it there exists the limit}

\begin{eqnarray}
\label{2.54}
& & \int d^{4n}x L_{c}^{- 1}
\bigl[ (\prod_{m = 1}^{l} x_{j_{m} }^{0} )^{- k} f
\bigr]_{x_{\underline{j} }^{0} }
(\underline{x} ) \prod_{i = 1}^{n} \phi_{i} (x_{i} ) =
\nonumber \\
& & \lim_{n_{1},...,n_{l} \rightarrow \infty , n_{i} \in {\bf Z} }
\lim_{T_{1},...,T_{l} \rightarrow - \infty , T_{i} \in {\bf R} }
\int d^{4n}x \bigl( \prod_{m = 1}^{l} x_{j_{m} }^{0} \bigr)^{- k}
f(\underline{x} ) \times \nonumber \\
& &\bigl( \prod_{i = 1, i \neq j_{1},...,j_{l} }^{n} \phi_{i} (x_{i})\bigr)
\prod_{m = 1}^{l} L_{c}^{- 1}[\phi_{j_{m}} ]_{x_{j_{m}}^{0}}
(x_{j_{m}};n_{m},T_{m}), \\
& & L_{c}^{- 1}[\phi ]_{x^{0} } (x;n,T) = (n!)^{- 1}
\bigl( \frac{\partial }{\partial x^{0} } \bigr)^{n + 1}
((x^{0})^{n} \theta (x^{0})  h_{T} (- x^{0}) \phi (n(x^{0})^{- 1},{\bf x})),
\nonumber
\end{eqnarray}
{\it where the function} $h_{T} (x^{0})$ {\it is given by the equality}
(\ref{2.13}).

{\it The limit} (\ref{2.54}) {\it defines the inversion formula
for the Laplace transformation:}

\begin{eqnarray}
\label{2.55}
& & ( f(\underline{x} ), \prod_{i = 1}^{n} \phi_{i} (x_{i})) =
\int d^{4n}x \bigl( ( \prod_{m = 1}^{l}
\frac{\partial }{\partial x_{j_{m}}^{0}})^{k}
L_{c}^{- 1}\bigl[ (\prod_{m = 1}^{l} x_{j_{m}}^{0})^{- k}
f\bigr]_{x_{\underline{j} }^{0}} (\underline{x} ) \bigr)
\times \nonumber \\
& & \int_{0}^{\infty } dy_{j_{1}}^{0}...\int_{0}^{\infty } dy_{j_{m}}^{0}
\exp \{ - \sum_{m = 1}^{l} x_{j_{m}}^{0}y_{j_{m}}^{0}\}
\bigl( \prod_{m = 1}^{l} \phi_{j_{m}} (y_{j_{m}}^{0},{\bf x}_{j_{m}})\bigr)
\prod_{i = 1, i \neq  j_{1},...,j_{l}}^{n} \phi_{i} (x_{i})
\end{eqnarray}
{\it for all functions} $\phi_{i} (x) \in S({\bf R}_{+}^{4} )$, $i = 1$,...,
$n$, {\it and for any integer} $k > K$.

\noindent {\it Proof.} Theorem from \cite[Section 3.8]{vla} implies
that a tempered distribution with support in the closure
$\overline{{\bf R}}_{+}^{4n} $ has the following form

\begin{equation}
\label{2.56}
g(\underline{x} ) = \sum_{|\underline{m} | \leq K - 1}
(\frac{\partial }{\partial \underline{y} })^{\underline{m} }
\mu_{\underline{m} } (\underline{y} ),
\end{equation}
where we use the standard multiindex notations and
$\mu_{\underline{m} } (\underline{y} )$ are the measures with tempered
growth and supports in $\overline{{\bf R}}_{+}^{4n} $. Similarly to the
proof of Corollary 2.6 the relation (\ref{2.56}) may be written as

\begin{equation}
\label{2.57}
g(\underline{x} ) = \bigl( \prod_{s = 1}^{l}
\frac{\partial }{\partial y_{j_{s}}^{0}} \bigr)^{K}
\sum_{|\underline{m} | \leq K - 1, m_{j_{1}}^{0} = ... = m_{j_{l}}^{0} = 0}
(\frac{\partial }{\partial \underline{y} })^{\underline{m} }
\nu_{\underline{m} } (\underline{y} ),
\end{equation}
where $\nu_{\underline{m} } (\underline{y} )$ are the measures with tempered
growth and supports in $\overline{{\bf R}}_{+}^{4n} $. Let us integrate
the distribution (\ref{2.57}) with the test function which is the
product of the funtions:
$\exp \{ - x_{j}^{0}y_{j}^{0}\} $$\phi_{j} ({\bf y}_{j})$,
$\phi_{j} ({\bf y}) \in S({\bf R}^{3})$, $j \in \{ j_{1},...,j_{l}\} $, and
$\exp \{ - x_{i}^{0}y_{i}^{0}\} $$\phi_{i} (x_{i}^{0},{\bf y}_{i})$,
$\phi_{i} (x) \in S({\bf R}_{+}^{4})$, $1 \leq i \leq n$,
$i \neq  j_{1},...,j_{l}$. Let us divide the obtained integral by
$(x_{j_{1}}^{0} \cdots x_{j_{l}}^{0})^{K}$. If the distribution
$f(\underline{x} )$ is the Laplace transform of the tempered
distribution (\ref{2.57}) with respect to time variables we get

\begin{eqnarray}
\label{2.58}
& & \bigl( \prod_{m = 1}^{l} x_{j_{m}}^{0} \bigr)^{- K} \int
\bigl( \prod_{i = 1, i \neq j_{1},...,j_{l}}^{n} dx_{i}^{0} \bigr)
d^{3n}{\bf x} f(\underline{x} )
\bigl( \prod_{m = 1}^{l} \phi_{j_{m}} ({\bf x}_{j_{m}})\bigr)
\prod_{i = 1, i \neq  j_{1},...,j_{l}}^{n} \phi_{i} (x_{i}) =
\nonumber \\
& & \int d\nu (y_{1},...,y_{l})
\exp \{ - \sum_{m = 1}^{l} x_{j_{m}}^{0}y_{m}\} ,
\end{eqnarray}
where the measure $\nu (y_{1},...,y_{l})$ with support in
$(\overline{{\bf R}}_{+} )^{\times l}$ is defined by

\begin{eqnarray}
\label{2.59}
& & \int d\nu (y_{1},...,y_{l}) \psi (y_{1},...,y_{l}) =
\sum_{|\underline{m} | \leq K - 1, m_{j_{1}}^{0} = ... = m_{j_{l}}^{0} = 0}
\int d\nu_{\underline{m} } (\underline{y} )
\psi (y_{j_{1}}^{0},...,y_{j_{l}}^{0}) \times \nonumber \\
& & (- \frac{\partial }{\partial \underline{y} })^{\underline{m} }
\bigl[ ( \prod_{m = 1}^{l} \phi_{j_{m}} ({\bf y}_{j_{m}}))
( \prod_{i = 1, i \neq  j_{1},...,j_{l}}^{n} \int dx_{i}^{0}
\exp \{ - x_{i}^{0}y_{i}^{0}\} \phi_{i} (x_{i}^{0},{\bf y}_{i}))
\bigr]
\end{eqnarray}
The measure $\nu (y_{1},...,y_{l})$ is the difference
$\nu_{1} (y_{1},...,y_{l}) - \nu_{2} (y_{1},...,y_{l})$  of two
positive measures with tempered growth and supports in
$(\overline{{\bf R}}_{+} )^{\times l}$. Then the left--hand side of
the equality (\ref{2.58}) is equal to

\begin{equation}
\label{2.60}
\int [d\nu_{1} (y_{1},...,y_{l}) - d\nu_{2} (y_{1},...,y_{l})]
\exp \{ - \sum_{m = 1}^{l} x_{j_{m}}^{0}y_{m}\} .
\end{equation}
The expression (\ref{2.60}), considered as the function on
$(\overline{{\bf R}}_{-} )^{\times l}$, is the difference
$f_{1}(x_{j_{1}}^{0},...,x_{j_{l}}^{0}) -$
$f_{2}(x_{j_{1}}^{0},...,x_{j_{l}}^{0})$ of two absolutely
monotonic with respect each variable distributions from
$S^{\prime }(({\bf R}_{-})^{\times l})$.

The arguments of the proof of Proposition 2.4 lead to the existence
of the limit (\ref{2.54}) for any test function which is the product
of the functions: $\psi_{j} (x_{j}^{0})\phi_{j} ({\bf x}_{j})$, where
$\psi_{j} (x^{0}) \in S({\bf R})$, $\phi_{j} ({\bf x}) \in S({\bf R}^{3})$,
$j =  j_{1},...,j_{l}$, and $\phi_{i} (x_{i}) \in S({\bf R}_{+}^{4})$,
$1 \leq i \leq n$, $i \neq j_{1},...,j_{l}$. Since the weak
convergence in $S^{\prime }$ implies the convergence in the topology
of the space $S^{\prime }$ (see \cite[Section 3.7]{vla}) the limit
is the polylinear functional continuous in each variable.
Now the nuclear theorem \cite[Chapter 1, Section 1, Theorem 6]{gvi}
implies the existence of the distribution
$L_{c}^{- 1}\bigl[ ( \prod_{m = 1}^{l} x_{j_{m}}^{0})^{- K}
f \bigr]_{x_{\underline{j} }^{0}} (\underline{x} )$ and the
equality (\ref{2.54}) for any test function which is the product
of the functions $\phi_{j} (x_{j}) \in S({\bf R}^{4})$,
$j = j_{1},...,j_{l}$, and $\phi_{i} (x_{i}) \in S({\bf R}_{+}^{4})$,
$1 \leq i \leq n$, $i \neq j_{1},...,j_{l}$. We use this set of the
functions in order to avoid the cumbersome notations. For the
special case $l = n$ the notations are simple:
$L_{c}^{- 1}\bigl[ ( \prod_{m = 1}^{l} x_{m}^{0})^{- K}
f \bigr]_{\underline{x}^{0} } (\underline{x} ) \in $
$S^{\prime }({\bf R}^{4n})$ and its support is in
$\overline{{\bf R}}_{+}^{4n} $.  The equality (\ref{2.54}) is valid
in this case for any test function
$\phi (\underline{x} ) \in S({\bf R}^{4n})$.

For the absolutely monotonic tempered distribution
$g(x) \in S^{\prime }({\bf R}_{-})$ the tempered distribution
$(- x)^{- m}g(x) \in S^{\prime }({\bf R}_{-})$ is absolutely
monotonic for any natural number $m$. It is possible therefore
to divide the expression (\ref{2.60}) by
$(x_{j_{1}}^{0} \cdots x_{j_{l}}^{0})^{k - K}$ and to prove the
above results for any integer $k > K$. For the integer $k > K$
Lemma 2.2 implies that the limits of type (\ref{2.7}) are equal to
zero. Then applying the arguments of the proof of Theorem 2.5 we
obtain the equality (\ref{2.55}).

\section{Exponentially convex distributions}
\setcounter{equation}{0}

\noindent Due to S.Bernstein \cite{ber} we call a tempered distribution
$f(x) \in S^{\prime }({\bf R}_{+})$ exponentially convex if it satisfies
the positivity condition (\ref{1.1}) for any function
$\phi (x) \in S({\bf R}_{+})$.

\noindent {\bf Proposition~3.1.} {\it For any exponentially convex
tempered  distribution} $f(x) \in S^{\prime }({\bf R}_{+})$
{\it the tempered distribution} $f(- x) \in S^{\prime }({\bf R}_{-})$
{\it is absolutely monotonic.}

\noindent {\it Proof.} Let us introduce the convolution function

\begin{equation}
\label{3.1}
\overline{\phi } \ast \phi (x) = \int_{- \infty }^{\infty }
\overline{\phi (x - y)} \phi (y) dy .
\end{equation}
For $\phi (x) \in S({\bf R}_{+})$ the function
$\overline{\phi } \ast \phi (x) \in S({\bf R}_{+})$. The condition
(\ref{1.1}) may be rewritten as
$(f(x),\overline{\phi } \ast \phi (x)) \geq 0$. Let
$F[\phi ](p)$ be the Fourier transform of the function
$\phi (x) \in S({\bf R}_{+})$. The definition (\ref{3.1}) implies
the following equality $F[\overline{\phi } \ast \phi ](p) = $
$\overline{F[\phi ](- p)} F[\phi ](p)$.

By $FS({\bf R}_{+})$ we denote the space of all analytic in the open
upper half plane and infinitely differentiable in the closed upper
half plane functions $\psi (z)$ such that the seminorms of the form

\begin{equation}
\label{3.2}
\sup_{{\rm Im}z \geq 0} (1 + |z|)^{n} |\frac{d^{m}\psi }{dz^{m}} (z)|
\end{equation}
are finite for all positive integers $m$ and $n$. The topology of
$FS({\bf R}_{+})$ is given by the set of the seminorms (\ref{3.2}).

Let us prove that the Fourier transformation defines an isomorphism
between two topological spaces: $S({\bf R}_{+})$ and $FS({\bf R}_{+})$.
Fourier transform $F[\phi ](x)$ of a function $\phi (x) \in S({\bf R}_{+})$
has an analytical continuation $F[\phi ](z)$ into the open upper half
plane. The function $F[\phi ](z)$ is infinitely differentiable in the
closed upper half plane. The inequality $y^{k}\exp \{ - yt\} \leq $
$C(k)t^{- k}$, valid for $t > 0$, $y \geq 0$ implies the following
estimation

\begin{eqnarray}
\label{3.3}
& & \sup_{x \in {\bf R}, y \geq 0} |x^{k_{1}}y^{k_{2}}
\frac{d^{m}F[\phi ]}{dz^{m}} (x + iy)| \leq \\
& & C \sup_{t > 0, 0 \leq l \leq k_{1}} (1 + t^{2})t^{- k_{2} - l}
|\frac{d^{k_{1} - l}}{dt^{k_{1} - l}} (t^{m}\phi (t))| . \nonumber
\end{eqnarray}
Therefore the Fourier transformation defines the continuous mapping
of the space $S({\bf R}_{+})$ into the space $FS({\bf R}_{+})$.
For a function $\psi (z) \in FS({\bf R}_{+})$ its restriction $\psi (x)$
on the real axis belongs to the Schwartz space $S({\bf R})$. A
straightforward application of Cauchy's theorem shows that the
inverse Fourier transform $F^{- 1}[\psi ](p)$ of the function $\psi (x)$
may be rewritten for $y > 0$ as

\begin{equation}
\label{3.4}
F^{- 1}[\psi ](p) = (2\pi )^{- 1}e^{py} \int_{- \infty }^{\infty }
\exp \{ - ipx\} \psi (x + iy) dx .
\end{equation}
Since any seminorm (\ref{3.2}) is finite we get $F^{- 1}[\psi ](p) = 0$
for $p < 0$ by tending $y$ in (\ref{3.4}) to infinity. Hence
$F^{- 1}[\psi ](p) \in S({\bf R}_{+})$. The inverse Fourier
transformation is the topological isomorphism of the Schwartz space
$S({\bf R})$. Any seminorm of the Schwartz space $S({\bf R})$ on the
subspace $FS({\bf R}_{+})$ is majorized by a corresponding seminorm of
the type (\ref{3.2}). Thus the inverse Fourier transformation is the
mapping of the space $FS({\bf R}_{+})$ into the space $S({\bf R}_{+})$.

For any natural number $k$ and for any number $\alpha > 0$ we define the
function

\begin{eqnarray}
\label{3.5}
& & F[\chi_{\alpha } ](z) = ((\alpha + z/i)^{1/2})^{k}
\exp \{ - \alpha (1 + z/i)^{1/2}\} ,\\
& & (a + z/i)^{1/2} = (x^{2} + (a + y)^{2})^{1/4}
\exp \{ - i/2 \arctan [x(a + y)^{- 1}]\} ,\nonumber
\end{eqnarray}
holomorphic in the open upper half plane. Due to the estimation
${\rm Re}(1 + z/i)^{1/2} \geq (|z|/2)^{1/2}$, valid in the closed
upper half plane, the function (\ref{3.5}) belongs to the space
$FS({\bf R}_{+})$. Hence its inverse Fourier transform
$\chi_{\alpha } (x)$ belongs to the space $S({\bf R}_{+})$,
as the function $\chi_{\alpha } (x - t)$
for any positive number $t$. Therefore the convolution function
$\overline{\chi }_{\alpha } \ast \chi_{\alpha } (x - 2t) = $
$\overline{\chi_{\alpha } (\bullet - t)} \ast \chi_{\alpha } (\bullet - t)(x)$
belongs to the space $S({\bf R}_{+})$ for any $t \geq 0$. Now the
positivity condition (\ref{1.1}) for the exponentially convex tempered
distribution $f(x) \in S^{\prime }({\bf R}_{+})$ implies the inequality
$(f(x),\overline{\chi }_{\alpha } \ast \chi_{\alpha } (x - 2t)) \geq 0$
for any $t \geq 0$. By integration of this inequality with positive
function $1/2 \phi (2t) \in S({\bf R}_{+})$ we obtain

\begin{equation}
\label{3.6}
(f(x),(\overline{\chi }_{\alpha } \ast \chi_{\alpha } ) \ast \phi (x)) \geq 0.
\end{equation}
In view of the definition (\ref{3.5}) we get
$\overline{F[\chi_{\alpha } ](- x)} = F[\chi_{\alpha } ](x)$. Now it
is easy to show that

\begin{equation}
\label{3.7}
(\overline{\chi }_{\alpha } \ast \chi_{\alpha } ) \ast \phi (x) =
(2\pi )^{- 1} \int_{- \infty }^{\infty } dp e^{- ipx} F[\phi ](p)
(F[\chi_{\alpha } ](p))^{2}
\end{equation}
When $\alpha \rightarrow + 0$ the functions
$F[\phi ](z)(F[\chi_{\alpha } ](z))^{2}$ converge to the function
$(z/i)^{k}F[\phi ](z)$ in the topology of the space $FS({\bf R}_{+})$.
Then the left--hand side of the equality (\ref{3.7}) converges to the
function $\frac{d^{k}\phi }{dx^{k}} (x)$ in the topology of the space
$S({\bf R}_{+})$ as $\alpha \rightarrow + 0$. This implies that the
inequality (\ref{3.6}) converges as $\alpha \rightarrow + 0$ to the
inequality

\begin{equation}
\label{3.8}
(- 1)^{k}(\frac{d^{k}f}{dx^{k}} (x),\phi (x)) \geq 0 .
\end{equation}
It follows from the inequalities (\ref{3.8}) for arbitrary natural
numbers $k$ and for arbitrary positive functions
$\phi (x) \in S({\bf R}_{+})$ that the tempered distribution
$f(- x) \in S^{\prime }({\bf R}_{-})$ is absolutely monotonic.

The distribution $f(x) \in S^{\prime }({\bf R}_{+}^{4})$ satisfies
the Osterwalder--Schrader positivity condition, if for any function
$\phi (x) \in S({\bf R}_{+}^{4})$

\begin{equation}
\label{3.9}
\int d^{4}x d^{4}y  f(x^{0} + y^{0},{\bf x} - {\bf y})
\overline{\phi (x)} \phi (y)  \geq 0 .
\end{equation}
Therefore the Osterwalder--Schrader positivity condition (\ref{3.9})
is the condition of exponentially convexity with respect to the time
variable and it is the condition of positively definiteness with
respect to the space variables. By using the proof of Proposition 3.1
and the proof of Theorem 1 from \cite[Chapter 2, Section 3.1]{gvi}
it is possible to show that the Fourier transform
$F_{{\bf x}}[f](x^{0},{\bf x})$ with respect to the space variables
of the distrirution $f(x) \in S^{\prime }({\bf R}_{+}^{4})$, satisfying
the condition (\ref{3.9}), satisfies the following condition

\begin{equation}
\label{3.10}
(- \frac{\partial }{\partial x^{0}} )^{n} F_{{\bf x}}[f](x^{0},{\bf x})
\geq 0
\end{equation}
for any natural number $n = 0,1$,... .

\noindent {\bf Lemma~3.2.} {\it Let tempered  distribution}
$f(x) \in S^{\prime }({\bf R}_{+}^{4})$ {\it satisfy the
Osterwalder--Schrader positivity condition} (\ref{3.9})
{\it then for any function} $\phi (x) \in S({\bf R}_{-}^{4})$

\begin{equation}
\label{3.11}
\lim_{t \rightarrow - \infty} (f(x),\phi (- x^{0} - t,{\bf x})) =
\int_{{\bf R^{3}}} L_{0}^{- 1}[f]_{x^{0}} ({\bf x}) d^{3}{\bf x}
\int_{- \infty }^{\infty } \phi (x^{0} ,{\bf x}) dx^{0},
\end{equation}
{\it where the tempered distribution} $L_{0}^{- 1}[f]_{x^{0}} ({\bf x})$
$\in S^{\prime }({\bf R}^{3})$ {\it is positively definite, and for any
natural number} $k = 1,2$,...

\begin{equation}
\label{3.12}
\lim_{t \rightarrow - \infty} t^{k}
((\frac{\partial }{\partial x^{0}} )^{k}f(x),\phi (- x^{0} - t,{\bf x})) = 0.
\end{equation}
{\it If the tempered distribution} $f(x) \in S^{\prime }({\bf R}_{+}^{4})$
{\it satisfies the Osterwalder--Schrader positivity condition} (\ref{3.9})
{\it the tempered distribution}
$(x^{0})^{- 1}f(x) \in S^{\prime }({\bf R}_{+}^{4})$ {\it satisfies the
inequalities} (\ref{3.10}) {\it and the limits} (\ref{3.11}), (\ref{3.12})
{\it for this distribution are equal to zero.}

\noindent {\it Proof.} Let the Fourier transform
$F_{{\bf x}}[\phi ](x^{0},{\bf x})$ with respect to the space variables of
a function $\phi (x) \in S({\bf R}_{-}^{4})$ be a positive function.
Then by the straightforward application of the inequalities (\ref{3.10})
and by the proof of Lemma 2.2 we can prove the relations (\ref{3.12}) and
the existence of the limit (\ref{3.11}).

Due to the Theorem 2 from \cite[Chapter 2, Section 2.2]{gvi} the
inequalities (\ref{3.10}) imply the following estimation for any
function $\phi (x) \in S({\bf R}_{-}^{4})$

\begin{equation}
\label{3.13}
|((\frac{\partial }{\partial x^{0}} )^{n}f(x),\phi (- x^{0},{\bf x}))| \leq
C\sup_{x^{0} \leq 0 , {\bf x} \in {\bf R}^{3}} (1 + |x|^{2})^{p}
|F_{{\bf x}}[\phi ](x^{0},{\bf x})|,
\end{equation}
where the numbers $C$ and $p$ depend on the natural number $n = 0,1,2$,... .
Let $\alpha ({\bf x})$ be an infinitely differentiable positive
function with a compact support and let it be equal to one into some
neighbourhood of zero. For a function $\phi (x) \in S({\bf R}_{-}^{4})$
we define $M = \sup |F_{{\bf x}}[\phi ](x)|$. The difference of two
positive functions from $S({\bf R}_{-}^{4})$:

\begin{eqnarray}
\label{3.14}
& & F_{{\bf x}}[\phi_{1m} ](x) = \exp \{ m^{- 1}(x^{0} + (x^{0})^{- 1})\}
\theta (- x^{0})\alpha (m^{- 1}{\bf x})M \\
& & F_{{\bf x}}[\phi_{2m} ](x) = \exp \{ m^{- 1}(x^{0} + (x^{0})^{- 1})\}
\theta (- x^{0})\alpha (m^{- 1}{\bf x})(M - F_{{\bf x}}[\phi ](x))\nonumber
\end{eqnarray}
converges as $m \rightarrow \infty $ to the function
$F_{{\bf x}}[\phi ](x) \in S({\bf R}_{-}^{4})$ in the norm (\ref{3.13}).
Due to the inequality (\ref{3.13}) it implies now the relations
(\ref{3.12}) and the existence of the limit (\ref{3.11}) for any
function $\phi (x) \in S({\bf R}_{-}^{4})$.

Let the integral of a positive function $h(x^{0}) \in S({\bf R}_{-})$ be
equal to one. Any function $\phi (x) \in S({\bf R}_{-}^{4})$ may be
represented as

\begin{equation}
\label{3.15}
\phi (x) = h(x^{0}) \int_{- \infty }^{\infty } \phi (y^{0} ,{\bf x}) dy^{0}
+ \frac{\partial \psi}{\partial x^{0}} (x),
\end{equation}
where the function $\psi (x) \in S({\bf R}_{-}^{4})$. The existence of
the limit (\ref{3.11}) and the equalities (\ref{3.15}) and (\ref{3.12})
for $k = 1$ provide the equality (\ref{3.11}) for any function
$\phi (x) \in S({\bf R}_{-}^{4})$. In virtue of the inequality (\ref{3.10})
for $n = 0$ the tempered distribution $L_{0}^{- 1}[f]_{x^{0}} ({\bf x})$
is positively definite. The proof of the last part of Lemma 3.2
follows the arguments of Lemma 2.2.

For any function $\phi (x) \in S({\bf R}_{-}^{4})$ and for any integer
$k = 1,2$,... , we define the function

\begin{equation}
\label{3.16}
\phi_{x^{0}}^{(- k)} (x) = - ((k - 1)!)^{- 1} \int_{x^{0}}^{\infty }
(x^{0} - y^{0})^{k - 1} \phi (y^{0},{\bf x}) dy^{0}
\end{equation}
and $\phi_{x^{0}}^{(0)} (x) =\phi (x)$. The definition (\ref{3.16}) is
quite similar to the definition (\ref{2.12}). The infinitely
differentiable function (\ref{3.16}) equals zero for $x^{0} > 0$.
It is easy to see that
$(\frac{\partial }{\partial x^{0}} )^{l} \phi_{x^{0}}^{(- k)} (x) =$
$\phi_{x^{0}}^{(l - k)} (x)$ for $l \leq k$. Let a positive function
$h_{T}(x^{0})$ be given by the relation (\ref{2.13}) for some
positive function $h(x^{0}) \in D({\bf R}_{-})$, having the integral
equal one.

\noindent {\bf Lemma~3.3.} {\it Let a tempered  distribution}
$f(x) \in S^{\prime }({\bf R}_{+}^{4})$ {\it satisfy the
Osterwalder--Schrader positivity condition} (\ref{3.9}).
{\it Then for any function} $\phi (x) \in S({\bf R}_{-}^{4})$
{\it and for any integer} $k = 1,2$,...

\begin{eqnarray}
\label{3.17}
& & \lim_{T \rightarrow - \infty }
((\frac{\partial }{\partial x^{0}})^{k} f(x),h_{T}(- x^{0})
\phi_{x^{0}}^{(- k)} (- x^{0},{\bf x})) = \\
& & (f(x),\phi (- x^{0},{\bf x})) -
\int_{{\bf R}^{3}} L_{0}^{- 1}[f]_{x^{0}} ({\bf x})\int_{- \infty }^{\infty }
\phi (x^{0} ,{\bf x}) dx^{0}, \nonumber
\end{eqnarray}
{\it where the positively definite tempered distribution}
$L_{0}^{- 1}[f]_{x^{0}} ({\bf x})$$\in S^{\prime }({\bf R}^{3})$
{\it is defined by the equality} (\ref{3.11}).

The proof of Lemma 3.3 is exactly analogous to that of Lemma 2.3
and can be omitted.

For a tempered distribution $f(x) \in S^{\prime }({\bf R}_{+}^{4})$
satisfying the Osterwalder--Schrader positivity condition (\ref{3.9})
we define a functional on the space $S({\bf R}^{4})$
by the following relation

\begin{eqnarray}
\label{3.18}
& &(L_{c}^{- 1}[f]_{x^{0}} (x;n,T),\phi (x)) =
(f(x),L_{c}^{- 1}[\phi ]_{x^{0}} (x;n,T)) = \\
& & (n!)^{- 1} ((x^{0})^{n}(- \frac{\partial }{\partial x^{0}})^{n + 1} f
(x^{0},{\bf x}),\theta (x^{0})h_{T}(- x^{0})\phi (n(x^{0})^{- 1} ,{\bf x}))
,\nonumber
\end{eqnarray}
where $n$ is a positive integer and the function $h_{T}(x^{0})$  is
given by the equality (\ref{2.13}). It is easy to prove that the
tempered distribution
$F_{{\bf x}}[L_{c}^{- 1}[f]_{x^{0}} (\bullet ;n,T)](x)$ is positive
and its support is in the closure of ${\bf R}_{+}^{4}$.

\noindent {\bf Proposition~3.4.} {\it Let a tempered  distribution}
$f(x) \in S^{\prime }({\bf R}_{+}^{4})$ {\it satisfy the
Osterwalder--Schrader positivity condition} (\ref{3.9}).
{\it Then in the topological space} $S^{\prime }({\bf R}^{4})$
{\it there exists the limit}

\begin{equation}
\label{3.19}
\lim_{n \rightarrow \infty } \lim_{T \rightarrow - \infty }
L_{c}^{- 1}[f]_{x^{0}} (x;n,T) = L_{c}^{- 1}[f]_{x^{0}} (x)
\end{equation}
{\it The tempered distribution}
$F_{{\bf x}}[L_{c}^{- 1}[f]_{x^{0}} (\bullet )](x)$
$\in S^{\prime }({\bf R}^{4})$ {\it is positive and its support is
in the closure of ${\bf R}_{+}^{4}$.}

\noindent {\it Proof.} The number $N(f)$ is defined by means of the
estimation similar to the estimation (\ref{2.3}). The lemma
analogous to Lemma 2.1 is valid for the tempered distribution
$f(x) \in S^{\prime }({\bf R}_{+}^{4})$. By using the inequalities
(\ref{3.10}) and Theorem 2 from \cite[Chapter 2, Section 2.2]{gvi}
it is possible to prove the estimation similar to (\ref{2.23})

\begin{equation}
\label{3.20}
|(L_{c}^{- 1}[f]_{x^{0}} (x;n,T),\phi (x))| \leq
C_{n,T}(N(f) + 1) \sup_{x^{0} \geq 0, {\bf x} \in {\bf R}^{3} }
(1 + x^{0})^{N(f) + 1} (1 + |{\bf x}|)^{p} |F_{{\bf x}}[\phi ](x)| .
\end{equation}
The arguments analogous to those of the proof of Proposition 2.4
allow us to replace the constant $C_{n,T}(N(f) + 1)$ by the
constant independent of the numbers $n$ and $T$.

The inequalities (\ref{3.10}) and Proposition 2.4 imply that the
limit (\ref{3.19}) exists on every test function from $S({\bf R}^{4})$
whose Fourier transform with respect to the space variables is a
positive function. Let $\alpha (x) \in D({\bf R}^{4})$ be positive
and let it be equal to one into some neighbourhood of zero. For a
function $\phi \in S({\bf R}^{4})$ we define
$M = \sup |F_{{\bf x}}[\phi ](x)|$. The difference of two positive
functions from $S({\bf R}^{4})$

\begin{eqnarray}
\label{3.21}
& & F_{{\bf x}}[\phi_{1m} ](x) = \alpha (m^{- 1}x) M ,\\
& & F_{{\bf x}}[\phi_{2m} ](x) = \alpha (m^{- 1}x)
(M - F_{{\bf x}}[\phi ](x)) \nonumber
\end{eqnarray}
converges as $m \rightarrow \infty $ to the function $F_{{\bf x}}[\phi ](x)$
$\in S({\bf R}^{4})$ in the norm (\ref{3.20}). Since the constant
$C_{n,T}(N(f) + 1)$ in the estimation (\ref{3.20}) may be replaced
by the constant independent of the numbers $n$ and $T$ it implies the
existence of the limit (\ref{3.19}) on every test function from
$S({\bf R}^{4})$. Therefore due to \cite[Section 3.7]{vla} the
sequence of tempered distributions $L_{c}^{- 1}[f]_{x^{0}} (x;n,T)$
converges in topology of the space $S^{\prime }({\bf R}^{4})$. It
follows now from the inequalities (\ref{3.10}) and the definition
(\ref{3.18}) that the tempered distribution
$F_{{\bf x}}[L_{c}^{- 1}[f]_{x^{0}} (\bullet )](x)$
$\in S^{\prime }({\bf R}^{4})$ is positive and its support is in
the closure of ${\bf R}_{+}^{4}$.

A straightforward application of the arguments of the proof of
Theorem 2.5 provides the following theorem.

\noindent {\bf Theorem~3.5.} {\it Let a tempered  distribution}
$f(x) \in S^{\prime }({\bf R}_{+}^{4})$ {\it satisfy the
Osterwalder--Schrader positivity condition} (\ref{3.9}).
{\it Then the following representation}

\begin{eqnarray}
\label{3.22}
& & (f(x),\phi (x)) =
\int_{{\bf R^{3}}} L_{0}^{- 1}[f]_{x^{0}} ({\bf x}) d^{3}{\bf x}
\int_{- \infty }^{\infty } \phi (p^{0} ,{\bf x})) dp^{0} + \\
& & \int_{{\bf R^{4}}} L_{c}^{- 1}[f]_{x^{0}} (x) d^{4}x
\int_{- \infty }^{\infty } \exp \{ - x^{0}p^{0}\}
\phi (p^{0} ,{\bf x})) dp^{0} \nonumber
\end{eqnarray}
{\it is valid for any function} $\phi (x) \in S({\bf R}_{+}^{4})$.
{\it Here the positively definite tempered distribution}
$L_{0}^{- 1}[f]_{x^{0}} ({\bf x})$$\in S^{\prime }({\bf R}^{3})$
{\it is defined by the equality} (\ref{3.11}) {\it and the tempered
distribution} $L_{c}^{- 1}[f]_{x^{0}} (x)$$\in S^{\prime }({\bf R}^{4})$
{\it defined by the relations} (\ref{3.18}), (\ref{3.19}).
{\it The distribution}
$F_{{\bf x}}[L_{c}^{- 1}$$[f]_{x^{0}} $$(\bullet )]$$(x)$
 $\in S^{\prime }({\bf R}^{4})$ {\it is the positive measure
with tempered growth and support in the closure of} ${\bf R}_{+}^{4}$.

\section{Revised Osterwalder--Schrader theorem}
\setcounter{equation}{0}

\noindent We deal with the theory of one Hermitian scalar field.
By using the below results and Chapter 6 of the paper \cite{os1}
it is possible to formulate the extended Osterwalder--Schrader
axioms and to prove the revised Osterwalder--Schrader theorem
for the theories of arbitrary spinor fields.

We introduce some notation from the papers \cite{os1} and \cite{os2}.
We define the following open sets in ${\bf R}^{4n}$:
${\bf R}_{<}^{4n} = $$\{ \underline{x} \in {\bf R}^{4n} |$
$ x_{j + 1}^{0} > x_{j}^{0}, j = 1,...,n - 1\}$ and
${\bf R}_{0}^{4n} = $$\{ \underline{x} \in {\bf R}^{4n} |$
$ x_{i} \neq x_{j}, 1 \leq i < j \leq n \}$. For $O$ an open set in
${\bf R}^{4n}$, the space $S(O)$ is defined above. On $S({\bf R}^{4n})$
we define two involutions

\begin{eqnarray}
\label{4.1}
& & f^{\ast }(x_{1},...,x_{n}) = \overline{f} (x_{n},...,x_{1})\\
& & \theta f(x_{1},...,x_{n}) = f(\theta x_{1},...,\theta x_{n}),
\nonumber
\end{eqnarray}
where $\theta x = (- x^{0},{\bf x})$  and $\overline{f}$ means
complex conjugation. The space $S({\bf R}_{<}^{4n})$ is invariant
under the involution $f \rightarrow \theta f^{\ast }$. Let
$f \in S({\bf R}^{4n})$, $R \in SO_{4}$ be an element in the
rotation group, $a \in {\bf R}^{4}$ and $\pi \in P_{n}$ be an
element in the group of all permutations of $n$ objects (the letter
$S_{n}$ will be used elsewhere). Then we define $f_{(a,R)}$ and
$f^{\pi }$ by $f_{(a,R)}(x_{1},...,x_{n}) = $$f(Rx_{1} + a,...,Rx_{n} + a)$
and $f^{\pi }(x_{1},...,x_{n}) = $$f(x_{\pi (1)},...,x_{\pi (n)})$.

We remind the Osterwalder--Schrader axioms \cite{os1} for the
Schwinger functions (Euclidean Green's functions). The set of the
Schwinger functions $\{ s_{n}\} $ is a sequence of distributions
$s_{n}(x_{1},...,x_{n})$ with the following properties

\noindent E0. {\it Distributions}

\noindent $s_{0} \equiv 1$, $s_{n} \in S^{\prime }({\bf R}_{0}^{4n})$
and

\begin{equation}
\label{4.2}
\overline{(s_{n},f)} = (s_{n},\theta f^{\ast })
\end{equation}
for all functions $f \in S({\bf R}_{<}^{4n})$.

\noindent E1. {\it Euclidean invariance}

\begin{equation}
\label{4.3}
(s_{n},f_{(a,R)}) = (s_{n},f)
\end{equation}
for all $R \in SO_{4}$, $a \in {\bf R}^{4}$ and
$f \in S({\bf R}_{0}^{4n})$.

\noindent E2. {\it Positivity}

\begin{equation}
\label{4.4}
\sum_{n,m} (s_{n + m},\theta f_{n}^{\ast }\otimes f_{m}) \geq 0
\end{equation}
for all finite sequences of the functions
$f_{n} \in S({\bf R}_{<}^{4n} \cap {\bf R}_{+}^{4n})$, where the
function $(f \otimes g)(x_{1},...,x_{n +m}) =$
$f(x_{1},...,x_{n})g(x_{n + 1},...,x_{n + m})$ is defined for all
functions $f \in S({\bf R}^{4n})$ and $g \in S({\bf R}^{4m})$.

\noindent E3. {\it Symmetry}

\begin{equation}
\label{4.5}
(s_{n},f^{\pi }) = (s_{n},f)
\end{equation}
for all permutations $\pi \in P_{n}$ and for all functions
$f \in S({\bf R}_{0}^{4n})$.

\noindent E4. {\it Cluster property}

\begin{equation}
\label{4.6}
\lim_{t \rightarrow \infty }
(s_{n + m},\theta f_{n}^{\ast }\otimes (g_{m})_{(ta,1)}) =
(s_{n},\theta f_{n}^{\ast })(s_{m},g_{m})
\end{equation}
for all $f_{n} \in S({\bf R}_{<}^{4n} \cap {\bf R}_{+}^{4n})$,
$g_{m} \in S({\bf R}_{<}^{4m} \cap {\bf R}_{+}^{4m})$,
$a = (0,{\bf a})$, ${\bf a} \in {\bf R}^{3}$.

Let us consider the restriction of the distribution
$s_{n} \in S^{\prime }({\bf R}_{0}^{4n})$ on the test functions
from the space $S({\bf R}_{<}^{4n})$. Then the translation
invariance (\ref{4.3}) implies

\begin{equation}
\label{4.7}
s_{n}(x_{1},...,x_{n}) = S_{n - 1}(x_{2} - x_{1},...,x_{n} - x_{n - 1}),
\end{equation}
where the distribution $S_{n - 1}(\underline{x} )$
$\in S^{\prime }({\bf R}_{+}^{4(n - 1)})$. We note that for the
function $g(x_{1},...,x_{n + 1}) =$
$f(x_{2} - x_{1},...,x_{n + 1} - x_{n})$ the definitions (\ref{4.1})
imply the equality $\theta g^{\ast }(x_{1},...,x_{n + 1}) =$
$(\theta_{p} f^{\ast })(x_{2} - x_{1},...,x_{n + 1} - x_{n})$,
where the involution

\begin{equation}
\label{4.8}
\theta_{p} f(x_{1},...,x_{n}) = f(- \theta x_{1},...,- \theta x_{n})
\end{equation}
leaves the space $S({\bf R}_{+}^{4n})$ invariant.

We substitute into the inequality (\ref{4.4}) the sequence consisting
of single function

\begin{equation}
\label{4.9}
f_{m + 1}(x_{1},...,x_{m + 1}) = \left\{ \begin{array}{ll}
\overline{\phi }_{1} (x_{1})\phi_{m} (x_{2} - x_{1},...,x_{m + 1} - x_{m}),
                             & m > 0 \\
\overline{\phi }_{1} (x_{1}),& m = 0,
\end{array}  \right.
\end{equation}
where the functions $\phi_{1} \in S({\bf R}_{+}^{4})$ and
$\phi_{m} \in S({\bf R}_{+}^{4m})$. Then by using the definitions
(\ref{4.7}) and (\ref{4.8}) we can rewrite the inequality (\ref{4.4})
for $m = 0$ in the form (\ref{3.9}) and for $m > 0$ in the
following form

\begin{equation}
\label{4.10}
\int d^{4}x d^{4}y S_{2m + 1} (\theta_{p} \phi_{m}^{\ast } ,x - \theta y,
\phi_{m} ) \overline{\phi }_{1} (x)\phi_{1} (y)  \geq 0.
\end{equation}
Here we introduce the distribution

\begin{equation}
\label{4.11}
\int d^{4}x S_{n + m + 1}(f_{n},x,f_{m}) f_{1}(x) =
\int d^{4(n + m + 1)}x S_{n + m + 1}(\underline{x} )
f_{n}\otimes f_{1}\otimes f_{m}(\underline{x} ),
\end{equation}
constructed from the distribution $S_{n + m + 1}(\underline{x} ) \in $
$S^{\prime }({\bf R}_{+}^{4(n + m + 1)})$ and the test functions
$f_{n} \in S({\bf R}_{+}^{4n})$, $f_{m} \in S({\bf R}_{+}^{4m})$.
For $n = 0$ or $m = 0$ the distribution (\ref{4.11}) is defined
in an obvious way.

The inequalities (\ref{4.10}) show that the distributions
$S_{2m + 1} (\theta_{p} \phi_{m}^{\ast } ,x,\phi_{m} )$ are
extremely significant. We formulate the new axiom exactly for
these distributions.

\noindent E5. {\it Weak spectral condition}

\noindent Let $S_{2m + 1}(\underline{x} ) \in $
$S^{\prime }({\bf R}_{+}^{4(2m + 1)})$, $m = 1,2$,... , be any
distribution defined by the relation (\ref{4.7}). Then there is
the natural number $K$ such that for any integers $k > K$,
$1 \leq l \leq m$ and for all test functions
$\psi_{i} (x) \in S({\bf R}^{4})$, $i = 1,...,l$,
$\psi_{j} (x) \in S({\bf R}_{+}^{4})$, $j = l + 1,...,m + 1$,
there exists the limit

\begin{eqnarray}
\label{4.12}
& & \lim_{n_{1},...,n_{l} \rightarrow \infty , n_{i} \in {\bf Z}}
\lim_{T_{1},...,T_{l} \rightarrow - \infty , T_{i} \in {\bf R}}
\int d^{4(2m + 1)}x
\bigl( \prod_{i = 1}^{l} x_{i}^{0}x_{2m + 2 - i}^{0}\bigr)^{- k}
S_{2m + 1}(\underline{x} ) \times \nonumber \\
& & [(L_{c}^{- 1}
[\psi_{1} \otimes \cdots \otimes \psi_{m} ]_{x_{1}^{0},...,x_{l}^{0}}
(\bullet ;\underline{n} ,\underline{T} )) \otimes \psi_{m + 1} \otimes
\nonumber \\
& & \theta_{p} (L_{c}^{- 1}
[\psi_{1} \otimes \cdots \otimes \psi_{m} ]_{x_{1}^{0},...,x_{l}^{0}}
(\bullet ;\underline{n} ,\underline{T} ))^{\ast }] (\underline{x} ),
\end{eqnarray}
where the function

\begin{equation}
\label{4.13}
L_{c}^{- 1}
[\psi_{1} \otimes \cdots \otimes \psi_{m} ]_{x_{1}^{0},...,x_{l}^{0}}
(\underline{x} ;\underline{n} ,\underline{T} )) =
\bigl( \prod_{i = 1}^{l} L_{c}^{- 1}[\psi_{i} ]_{x_{i}^{0}}
(x_{i};n_{i},T_{i})\bigr)
\bigl( \prod_{i = l + 1}^{m} \psi_{i} (x_{i})\bigr)
\end{equation}
and the function $L_{c}^{- 1}[\psi ]_{x^{0}}(x;n,T)$ is given by
the equality (\ref{2.54}).

Theorem 2.7 clarifies this weak spectral condition.

Let us remind the Wightman axioms \cite{swi} for the Wightman
distributions. The set of the Wightman distributions $\{ w_{n}\} $
is a sequence of distributions with the following properties

\noindent R0. {\it Temperedness}

\noindent $w_{0} \equiv 1$, $w_{n} \in S^{\prime }({\bf R}^{4n})$ and

\begin{equation}
\label{4.14}
\overline{(w_{n},f)}  = (w_{n},f^{\ast })
\end{equation}
for all $f \in S({\bf R}^{4n})$.

\noindent R1. {\it Relativistic invariance}

\begin{equation}
\label{4.15}
(w_{n},f_{(a,\Lambda )}) = (w_{n},f)
\end{equation}
for all vectors $a \in {\bf R}^{4}$, for all Lorentz transformations
$\Lambda \in L_{+}^{\uparrow } $ and for all functions
$f \in S({\bf R}^{4n})$, where the function
$f_{(a,\Lambda )}(\underline{x} ) =$
$f(\Lambda^{- 1} (x_{1} - a),...,\Lambda^{- 1} (x_{n} - a))$.

\noindent R2. {\it Positivity}

\begin{equation}
\label{4.16}
\sum_{n,m} (w_{n + m}, f_{n}^{\ast }\otimes f_{m}) \geq 0
\end{equation}
for all finite sequences of the functions $f_{n} \in S({\bf R}^{4n})$.

\noindent R3. {\it Local commutativity}

\noindent For all natural numbers $n > 0$ and $j = 1,...,n - 1$

\begin{equation}
\label{4.17}
w_{n}(x_{1},...,x_{j + 1},x_{j},..,x_{n}) =
w_{n}(x_{1},...,x_{j},x_{j + 1},..,x_{n})
\end{equation}
if the vector $x_{j + 1} - x_{j} \in {\bf R}^{4}$ is spacelike:
$(x_{j + 1} - x_{j},x_{j + 1} - x_{j}) \equiv $
$(x_{j + 1}^{0} - x_{j}^{0})^{2} - $
$\sum_{i = 1}^{3} (x_{j + 1}^{i} - x_{j}^{i})^{2}$$ < 0$.

\noindent R4. {\it Cluster property}

\begin{eqnarray}
\label{4.18}
& & \lim_{\lambda \rightarrow \infty }
w_{n + m}(x_{1},...,x_{n},x_{n + 1} +\lambda a,...,x_{n + m} +\lambda a) = \\
& & w_{n}(x_{1},...,x_{n})w_{m}(x_{n + 1},...,x_{n + m}) \nonumber
\end{eqnarray}
for all natural numbers $n,m > 0$ and for all spacelike vectors
$a \in {\bf R}^{4}$.

\noindent R5. {\it Spectral condition}

\noindent For all natural numbers $n > 1$ there exists the tempered
distribution $W_{n - 1}^{\sim } \in S^{\prime }({\bf R}^{4(n - 1)})$
with support in $\overline{V}_{+}^{\times n} $, where
$\overline{V}_{+}$ is the closed forward light cone, such that

\begin{equation}
\label{4.19}
w_{n}(\underline{x} ) = \int d^{4(n - 1)}p W_{n - 1}^{\sim }(\underline{p} )
\exp \{ i\sum_{j = 1}^{n - 1} (p_{j},(x_{j + 1} - x_{j}))\} .
\end{equation}

Now we are able to formulate the revised Osterwalder--Schrader theorem.

\noindent {\bf Theorem 4.1.} {\it To a given sequence of Wightman
distributions satisfying R0--R5, there corresponds a unique sequence
of Schwinger functions with the properties E0--E5. To a given sequence
of Schwinger functions satisfying E0--E5, there corresponds a unique
sequence of Wightman distributions with the properties R0--R5.}

\noindent {\it Proof.} We start from a relativistic field theory
given by a sequence of Wightman distributions, satisfying the axioms
R0--R5. Due to Theorem 3.5 from \cite{swi} the Wightman distribution
$w_{n}$ is the boundary value of the Wightman function
$w_{n}(z_{1},...,z_{n}) =$ $W_{n - 1}(z_{2} - z_{1},...,z_{n} - z_{n - 1})$,
where the function $W_{n - 1}(z_{1},...,z_{n - 1})$ is analytic in the
tube $T_{n - 1} = \{ z_{1},...,z_{n - 1}| $
${\rm Im}z_{i} \in V_{+}, i = 1,...,n - 1\} $. The Wightman function
$w_{n}(z_{1},...,z_{n})$ is Lorentz invariant (Lorentz covariant for
the theories of arbitrary spinor fields). The Bargmann Hall Wightman
theorem \cite[Theorem 2.11]{swi} implies that the function
$W_{n}(z_{1},...,z_{n})$ allows a single valued $L_{+}({\bf C})$
invariant ($L_{+}({\bf C})$ covariant for the theories of arbitrary
spinor fields) analytic continuation into the extended tube
$T_{n}^{\prime } = \cup_{A \in L_{+}({\bf C})} AT_{n}$.
By using Theorem 3.6 from \cite{swi} we conclude that the function
$w_{n}(z_{1},...,z_{n})$ has an $L_{+}({\bf C})$ invariant, single
valued, symmetric under the permutations analytic continuation into
the domain $IT_{n}^{p} = $$\{ z_{1},...,z_{n} |$
$(z_{\pi (2)} - z_{\pi (1)},...,z_{\pi (n)} - z_{\pi (n - 1)}) \in $
$T_{n - 1}^{\prime }$ for some permutation $\pi (1),...,\pi (n) $
of the numbers $1,...,n\} $. (For the theories of arbitrary spinor
fields this function has an $L_{+}({\bf C})$ covariant, single
valued analytic continuation into the domain $IT_{n}^{p}$ with
obvious symmetry properties under the permutations.) The set
$IT_{n}^{p}$ contains the set of the Euclidean points
$E_{n} = \{ z_{1},...,z_{n} |$$ {\rm Re}z_{k}^{0} = 0,$
$ {\rm Im}{\bf z}_{k} = 0, z_{k} \neq z_{j} $for all
$1 \leq k,j \leq n, k \neq j \} $. The restriction of the Wightman
functions to Euclidean points defines the Schwinger functions

\begin{equation}
\label{4.20}
s_{n}(x_{1},...,x_{n}) =
w_{n}((ix_{1}^{0},{\bf x}_{1}),...,(ix_{n}^{0},{\bf x}_{n})).
\end{equation}
The derivation of the extended Osterwalder--Schrader axioms
E0--E5 from the Wightman axioms follows the arguments of the
paper \cite{os1} and of Theorem 2.7.

Let $\{ s_{n}\} $ be a sequence of distributions satisfying the
extended Osterwalder--Schrader axioms E0--E5. If we substitute into
the inequality (\ref{4.4}) the sequence consisting of single
function (\ref{4.9}) for $m = 0$ we get the inequality (\ref{3.9})
for the distribution $S_{1}(x)$. Due to Theorem 3.5 this
distribution is the Laplace transform with respect to the time variable
of a tempered distribution with support in the closure
$\overline{{\bf R}}_{+}^{4} $. Let us substitute into the inequality
(\ref{4.4}) the sequence consisting of two functions
$f_{n + 1}(x)$ and $f_{m + 1}(x)$ of type (\ref{4.9}) with the
same function $\phi_{1} (x)$. Then we obtain the following inequality

\begin{equation}
\label{4.21}
\int d^{4}x d^{4}y S\{ \phi_{n} ,\phi_{m} \} (x - \theta y)
\overline{\phi_{1} (x)} \phi_{1} (y)  \geq 0,
\end{equation}
where the distribution

\begin{eqnarray}
\label{4.22}
& & S\{ \phi_{n} ,\phi_{m} \} (x) = S\{ \phi_{m} ,\phi_{n} \} (x) =
S_{2m + 1} (\theta_{p} \phi_{m}^{\ast } ,x, \phi_{m} ) +
S_{2n + 1} (\theta_{p} \phi_{n}^{\ast } ,x, \phi_{n} ) + \nonumber \\
& & S_{m + n + 1} (\theta_{p} \phi_{m}^{\ast } ,x, \phi_{n} ) +
S_{m + n + 1} (\theta_{p} \phi_{n}^{\ast } ,x, \phi_{m} ) .
\end{eqnarray}
This definition may be easily modified for the case $n = 0$ or $m = 0$

\begin{eqnarray}
\label{4.23}
& & S\{ \lambda ,\phi_{m}  \} (x) = S\{ \phi_{m} ,\lambda \} (x) = \\
& & S_{2m + 1} (\theta_{p} \phi_{m}^{\ast } ,x, \phi_{m} ) +
|\lambda |^{2} S_{1}(x) +
\lambda S_{m + 1} (\theta_{p} \phi_{m}^{\ast } ,x) +
\overline{\lambda } S_{m + 1} (x, \phi_{m} ), \nonumber
\end{eqnarray}
where $\lambda $ is a complex number. The equality (\ref{4.22})
implies the relation $S\{ \phi_{n} ,\phi_{n} \} (x) = $
$4S_{2n + 1} (\theta_{p} \phi_{n}^{\ast } ,x, \phi_{n} )$.
Hence the inequality (\ref{4.10}) is the particular case of the
inequality (\ref{4.21}) for $m =n$. It follows from the
definitions (\ref{4.22}), (\ref{4.23}) that the distribution
(\ref{4.11}) is the linear combination of the distributions
(\ref{4.22})  and (\ref{4.23})

\begin{eqnarray}
\label{4.24}
& & S_{m + n + 1} (\phi_{m} ,x, \phi_{n} ) =
1/2 S\{ \phi_{n} ,\theta_{p} \phi_{m}^{\ast } \} (x) +
i/2 S\{ \phi_{n} ,i\theta_{p} \phi_{m}^{\ast } \} (x) - \nonumber \\
& & (1 + i)/2 S_{2m + 1} (\phi_{m} ,x,\theta_{p} \phi_{m}^{\ast } ) -
(1 + i)/2 S_{2n + 1} (\theta_{p} \phi_{n}^{\ast } ,x,\phi_{n} ).
\end{eqnarray}
In particular for $m = 0$ or $n = 0$ and $\phi_{0} = 1$ we get

\begin{eqnarray}
\label{4.25}
& & S_{n + 1} (x, \phi_{n} ) =
1/2 S\{ 1,\phi_{n} \} (x) + i/2 S\{ i,\phi_{n} \} (x) - \nonumber \\
& & (1 + i)/2 S_{2n + 1} (\theta_{p} \phi_{n}^{\ast } ,x,\phi_{n} ) -
(1 + i)/2 S_{1}(x) ,
\end{eqnarray}

\begin{eqnarray}
\label{4.26}
& & S_{n + 1} (\phi_{n} ,x) =
1/2 S\{ 1,\theta_{p} \phi_{n}^{\ast } \} (x) +
i/2 S\{ 1,i\theta_{p} \phi_{n}^{\ast } \} (x) - \nonumber \\
& & (1 + i)/2 S_{2n + 1} (\phi_{n} ,x,\theta_{p} \phi_{n}^{\ast } ) -
(1 + i)/2 S_{1}(x) .
\end{eqnarray}
The inequalities (\ref{4.21}) imply that for any function
$\phi_{n} \in S({\bf R}_{+}^{4})$ every of four distributions,
depending on the variable $x$, in the right--hand side of the
equality (\ref{4.25}) is proportional to the distribution from
$S^{\prime }({\bf R}_{+}^{4})$, satisfying the Osterwalder--Schrader
positivity condition (\ref{3.9}). Due to Lemma 3.2 the limits
(\ref{3.11}) and (\ref{3.12}) are equal to zero for the
distribution $(x^{0})^{- k}S_{n + 1}(x,\phi_{n} ) \in $
$S^{\prime }({\bf R}_{+}^{4})$ if the integer $k > 0$.
It follows from Proposition 3.4 that for the distribution
$S_{n + 1}(\underline{x} ) \in $$S^{\prime }({\bf R}_{+}^{4(n + 1)})$
there exists the limit (\ref{2.54})  for the integers $l = 1$,
$j_{1} = 1$, $K = 0$. By the definition the support of the
distribution $L_{c}^{- 1}[(x_{1}^{0})^{- k}S_{n + 1}]_{x_{1}^{0}}$
$(x_{1},...,x_{n + 1})$ with respect to the first variable $x_{1}$
is in the closure $\overline{{\bf R}}_{+}^{4} $. Theorem 3.5
and Lemma 3.2 imply that for all functions
$\psi_{i} \in S({\bf R}_{+}^{4})$, $i = 1,...,n + 1$, and for any
integer $k > 0$ the following relation holds

\begin{eqnarray}
\label{4.27}
& & \int d^{4(n + 1)}x S_{n + 1}(\underline{x} )
\prod_{i = 1}^{n + 1} \psi_{i} (x_{i}) =
\int d^{4(n + 1)}x (\frac{\partial }{\partial x_{1}^{0}} )^{k}
L_{c}^{- 1}[(x_{1}^{0})^{- k}S_{n + 1}]_{x_{1}^{0}} (\underline{x} )
\times \nonumber \\
& & \int dy_{1}^{0} \exp \{ - x_{1}^{0}y_{1}^{0}\}
\psi_{1} (y_{1}^{0},{\bf x}_{1}) \prod_{i = 2}^{n + 1} \psi_{i} (x_{i})
\end{eqnarray}
In view of the equality (\ref{4.26}) all above results are
valid for the distribution $S_{n + 1} (\phi_{n} ,x) \in $
$S^{\prime }({\bf R}_{+}^{4})$, where the function $\phi_{n} \in$
$S({\bf R}_{+}^{4n})$.

The weak spectral condition E5 implies the existence of the limit

\begin{eqnarray}
\label{4.28}
& & \lim_{m \rightarrow \infty } \lim_{T \rightarrow - \infty }
\int d^{4(2m + 1)}x S_{2n + 1}(\underline{x} ) (x_{1}^{0})^{- k}
L_{c}^{- 1}[\psi_{1} ]_{x_{1}^{0}}(x_{1};m,T) \times \\
& & \bigl( \prod_{i = 2}^{n + 1} \psi_{i} (x_{i})\bigr)
\theta_{p} \bigl( \bigl( \prod_{i = 2}^{n} \overline{\psi}_{i}
(x_{2n + 2 - i})\bigr) (x_{2n + 1}^{0})^{- k}
L_{c}^{- 1}[\overline{\psi}_{1} ]_{x_{2n + 1}^{0}}(x_{2n + 1};m,T)\bigr)
\nonumber
\end{eqnarray}
for some positive integer $k$ and for all functions $\psi_{1} (x) \in$
$S({\bf R}^{4})$, $\psi_{i} (x) \in$ $S({\bf R}_{+}^{4})$,
$i = 2,...,n + 1$. Here the function
$L_{c}^{- 1}[\psi_{1} ]_{x^{0}}(x;m,T)$ is defined by the
relation (\ref{2.54}). The linear functional (\ref{4.28}) with
respect to the function $\psi_{n + 1} (x) \in$ $S({\bf R}_{+}^{4})$
is the tempered distribution from the space $S^{\prime }({\bf R}_{+}^{4})$.
It satisfies the Osterwalder--Schrader positivity condition
(\ref{3.9}). The similar arguments may be applied for the
distributions $S\{ 1,\theta_{p} \bigl( \prod_{i = 1}^{n} $
$\psi_{i} (x_{i})\bigr)^{\ast } \} (x)$ and
$S\{ 1,i\theta_{p} \bigl( \prod_{i = 1}^{n} $
$\psi_{i} (x_{i})\bigr)^{\ast } \} (x)$ in the right--hand side of the
equality of type (\ref{4.26}). Thus the limit (\ref{2.54})
$\bigl( L_{c}^{- 1}[(x_{1}^{0})^{- k}S_{n + 1}]_{x_{1}^{0}}(\underline{x} ),$
$\bigl( \prod_{i = 1}^{n + 1} \psi_{i} (x_{i})\bigr) \bigr) $,
the existence of which is proved above, for some positive
integer $k$ and for all functions $\psi_{1} (x) \in$
$S({\bf R}^{4})$, $\psi_{i} (x) \in$ $S({\bf R}_{+}^{4})$,
$i = 2,...,n + 1$, has the decomposition of type (\ref{4.26})
into four distributions with respect to the function
$\psi_{n + 1} (x) \in$ $S({\bf R}_{+}^{4})$. These distributions
are proportional to the distributions from $S^{\prime }({\bf R}_{+}^{4})$
satisfying the Osterwalder--Schrader positivity condition
(\ref{3.9}). Now Proposition 3.4 implies that for the
distribution $S_{n + 1}(\underline{x} ) \in $
$S^{\prime }({\bf R}_{+}^{4(n + 1)})$ there exists the limit
(\ref{2.54}) for $l = 2$, $j_{1} = 1$, $j_{2} = n + 1$ and
for some positive integer $k$. Due to the definition the
supports of this limiting distribution $L_{c}^{- 1}$
$[(x_{1}^{0}x_{n + 1}^{0})^{- k}S_{n + 1}]_{x_{1}^{0},x_{n + 1}^{0}}$
$(\underline{x} )$ with respect to the first and the last
variables are in the closure $\overline{{\bf R}}_{+}^{4} $.
Theorem 3.5, Lemma 3.2 and the relation (\ref{4.27}) imply
that for sufficiently large positive integer $k$ and for all
functions $\psi_{i} (x) \in$ $S({\bf R}_{+}^{4})$,
$i = 1,...,n + 1$, the following relation holds

\begin{eqnarray}
\label{4.29}
& \int & d^{4(n + 1)}x S_{n + 1}(\underline{x} )
         \prod_{i = 1}^{n + 1} \psi_{i} (x_{i}) =
  \int d^{4(n + 1)}x
      (\frac{\partial^{2} }{\partial x_{1}^{0} \partial x_{n + 1}^{0}} )^{k}
         L_{c}^{- 1}
         [(x_{1}^{0}x_{n + 1}^{0})^{- k}S_{n + 1}]_{x_{1}^{0},x_{n + 1}^{0}}
         (\underline{x} ) \times \nonumber \\
& \int & dy_{1}^{0}dy_{n + 1}^{0}
         \exp \{ - \sum_{i = 1,n + 1} x_{i}^{0}y_{i}^{0}\}
         \bigl( \prod_{i = 1,n + 1} \psi_{i} (y_{i}^{0},{\bf x}_{i}) \bigr)
         \prod_{i = 2}^{n} \psi_{i} (x_{i}) .
\end{eqnarray}

By using the weak spectral condition E5 and the equalities
(\ref{4.24}) it is possible to prove step by step that there
exists the limit (\ref{2.54}) for the distribution
$S_{n + 1}(\underline{x} ) \in $$S^{\prime }({\bf R}_{+}^{4(n + 1)})$,
for $l = n + 1$ and for some positive integer $k$. By the
definition the supports of this limiting distribution $L_{c}^{- 1}$
$[(\prod_{i = 1}^{n + 1} x_{i}^{0})^{- k}S_{n + 1}]_{\underline{x}^{0} }$
$(\underline{x} )$ with respect to any variable is in the closure
$\overline{{\bf R}}_{+}^{4} $. Since the weak convergence in the
space $S^{\prime }$ implies the convergence in the topology of the
space $S^{\prime }$ (see \cite[Section 3.7]{vla} ) the limit
$\bigl( L_{c}^{- 1}$
$[(\prod_{i = 1}^{n} x_{i}^{0})^{- k}S_{n + 1}]_{\underline{x}^{0} }$
$(\underline{x} )$,
$\bigl( \prod_{i = 1}^{n + 1} \psi_{i} (x_{i})\bigr) \bigr) $
is continuous in each function $\psi_{i} (x) \in$ $S({\bf R}^{4})$.
Hence the nuclear theorem \cite[Chapter 1, Section 1, Theorem 6]{gvi}
implies that $L_{c}^{- 1}$
$[(\prod_{i = 1}^{n + 1} x_{i}^{0})^{- k}S_{n + 1}]_{\underline{x}^{0} }$
$(\underline{x} ) \in $$S^{\prime }({\bf R}^{4(n + 1)})$. Its
support is in the closure $\overline{{\bf R}}_{+}^{4(n + 1)} $.
The application step by step of Theorem 3.5 and Lemma 3.2 gives
for sufficiently large positive integer $k$ and for all
functions $\psi_{i} (x) \in$ $S({\bf R}_{+}^{4})$, $i = 1,...,n + 1$
the following relation

\begin{eqnarray}
\label{4.30}
& \int & d^{4(n + 1)}x S_{n + 1}(\underline{x} )
\prod_{i = 1}^{n + 1} \psi_{i} (x_{i}) =
 \int d^{4(n + 1)}x
(\frac{\partial^{n + 1} }{\partial x_{1}^{0} \cdots \partial x_{n + 1}^{0}}
)^{k} L_{c}^{- 1}
[(\prod_{i = 1}^{n + 1} x_{i}^{0})^{- k}S_{n + 1}]_{\underline{x}^{0} }
(\underline{x} ) \times \nonumber  \\
& \int & dy_{1}^{0} \cdots dy_{n + 1}^{0}
\exp \{ - \sum_{i = 1}^{n + 1} x_{i}^{0}y_{i}^{0}\}
\bigl( \prod_{i = 1}^{n + 1} \psi_{i} (y_{i}^{0},{\bf x}_{i}) \bigr) .
\end{eqnarray}
Therefore for any $n = 1,2$,... the distribution
$S_{n}(\underline{x} ) \in$ $S^{\prime }({\bf R}_{+}^{4n})$ is the Laplace
transform of the tempered distribution from $S^{\prime }({\bf R}^{4n})$
with support in the closure $\overline{{\bf R}}_{+}^{4n} $.

Now the derivation of the Wightman axioms R0--R5 from the
Osterwalder--Schrader axioms E0--E4 follows the arguments of the
paper \cite{os1} .


\begin{thebibliography}{10}

\bibitem{os1}
Osterwalder, K., Schrader, R. : Axioms for Euclidean Green's Functions.
Commun. Math. Phys. {\bf 31}, 83--112 (1973)

\bibitem{os2}
Osterwalder, K., Schrader, R. : Axioms for Euclidean Green's Functions
2.  Commun. Math. Phys. {\bf 42}, 281--305 (1975)

\bibitem{gla}
Glaser, V. : On the Equivalence of the Euclidean and Wightman Formulation
of Field Theory. Commun. Math. Phys. {\bf 37}, 257--272 (1974)

\bibitem{ber}
Bernstein, S. : Sur les fonctions absolument monotones. Acta Math.
{\bf 52}, 1--66 (1929)

\bibitem{zin}
Zinoviev, Yu.M. : Inversion formulas for the Laplace transform. 1. The
octant. Teor. Mat. Fiz. {\bf 65}, 16--23 (1985); English transl. in
Theor. Math. Phys. {\bf 65}, (1985)

\bibitem{vla}
Vladimirov, V.S. : Methods of Theory of Functions of Many Complex
Variables. Cambridge MA : MIT Press 1966

\bibitem{gvi}
Gel'fand, I.M., Vilenkin, N.Ya. : Generalized Functions, Vol. 4.
New York : Academic Press 1964

\bibitem{swi}
Streater, R.F., Wightman, A.S. : PCT, spin and statistics and all that.
New York, Amsterdam : Benjamin 1964

\end{thebibliography}
\end{document}